\documentclass[sn-mathphys,Numbered]{sn-jnl}% Math and Physical Sciences Reference Style
\usepackage{graphicx}%
\usepackage{multirow}%
\usepackage{amsmath,amssymb,amsfonts}%
\usepackage{amsthm}%
\usepackage{mathrsfs}%
\usepackage[title]{appendix}%
\usepackage{xcolor}%
\usepackage{textcomp}%
\usepackage{manyfoot}%
\usepackage{booktabs}%
\usepackage{algorithm}%
\usepackage{algorithmicx}%
\usepackage{algpseudocode}%
\usepackage{listings}%
\usepackage{makecell}

\newcommand{\overbar}[1]{\mkern 1.5mu\overline{\mkern-1.5mu#1\mkern-1.5mu}\mkern 1.5mu}

\newcommand{\etal}{\textit{et al}.}
\newcommand{\ie}{\textit{i}.\textit{e}.}
\newcommand{\eg}{\textit{e}.\textit{g}.} 

\theoremstyle{thmstyleone}%
\theoremstyle{thmstyletwo}%

\theoremstyle{thmstylethree}%

\begin{document}

\title[Rethinking Skip Connections in Spiking Neural Networks]{Rethinking Skip Connections in Spiking Neural Networks with Time-To-First-Spike Coding} 

\author*[1]{\fnm{Youngeun} \sur{Kim}}\email{youngeun.kim@yale.edu}

\author[2]{\fnm{Adar} \sur{Kahana}}\email{adar$\_$kahana@brown.edu}

\author[1]{\fnm{Ruokai} \sur{Yin}}\email{ruokai.yin@yale.edu}

\author[1]{\fnm{Yuhang} \sur{Li}}\email{yuhang.li@yale.edu}

\author[2,3]{\fnm{Panos} \sur{Stinis}}\email{panagiotis.stinis@pnnl.gov}

\author[2,3]{\fnm{George Em} \sur{Karniadakis}}\email{george$\_$karniadakis@brown.edu}

\author[1]{\fnm{Priyadarshini} \sur{Panda}}\email{priya.panda@yale.edu}


\affil*[1]{Department of Electrical Engineering, Yale University, New Haven, CT, United States}

\affil[2]{Division of Applied Mathematics, Brown University, Providence, RI, United States}

\affil[3]{Advanced Computing, Mathematics and Data Division, Pacific Northwest National Laboratory, Richland, WA, United States}


\abstract{Time-To-First-Spike (TTFS) coding in Spiking Neural Networks (SNNs) offers significant advantages in terms of energy efficiency, closely mimicking the behavior of biological neurons. 
In this work, we delve into the role of skip connections, a widely used concept in Artificial Neural Networks (ANNs), within the domain of SNNs with TTFS coding. Our focus is on two distinct types of skip connection architectures: (1) addition-based skip connections, and (2) concatenation-based skip connections.
We find that addition-based skip connections introduce an additional delay in terms of spike timing. On the other hand, concatenation-based skip connections circumvent this delay but produce time gaps between after-convolution and skip connection paths, thereby restricting the effective mixing of information from these two paths. To mitigate these issues, we propose a novel approach involving a learnable delay for skip connections in the concatenation-based skip connection architecture. This approach successfully bridges the time gap between the convolutional and skip branches, facilitating improved information mixing.
We conduct experiments on public datasets including MNIST and Fashion-MNIST, illustrating the advantage of the skip connection in TTFS coding architectures. Additionally, we demonstrate the applicability of TTFS coding on beyond image recognition tasks and extend it to scientific machine-learning tasks, broadening the potential uses of SNNs.}




\maketitle

\section{Introduction}

The communication between spiking neurons in the brain, characterized by its binary, event-driven, and sparse nature, offers significant potential for creating flexible and energy-efficient artificial intelligence (AI) systems \citep{roy2019towards,christensen20222022}. 
Spiking Neural Networks (SNNs), unlike traditional Artificial Neural Networks (ANNs), leverage binary spikes, thereby offering a unique dimension of time in their operation.
Recent studies have shown promising results with SNNs, making them suitable for competitive and energy-efficient applications in neuromorphic hardware \citep{roy2019towards,panda2020toward,cao2015spiking,diehl2015unsupervised,comsa2020temporal}. 

A primary application of SNNs lies in image recognition \citep{roy2019towards,christensen20222022}. In order to transform a static image into binary spike trains, a range of coding schemes have been introduced \citep{comsa2020temporal,park2019fast,guo2021neural}. Rate coding conveys information through the firing rate of spikes \citep{wu2019direct,zheng2020going,zhang2020temporal,fang2020incorporating,lee2020enabling}. Phase coding, meanwhile, embeds temporal information in spike patterns utilizing a global oscillator \citep{montemurro2008phase}. In contrast, burst coding transmits spike bursts within brief time periods, which boosts the reliability of synaptic communication between neurons \citep{park2019fast}. While these coding schemes have proven successful in training SNNs, they generate a large number of spikes, which presents challenges when applied to ultra-low power devices.

To leverage temporal spike information in ultra-low power environments, researchers have increasingly focused on Time-To-First-Spike (TTFS) coding \citep{zhang2019tdsnn,rueckauer2018conversion}. The core concept involves representing information through spike timing, with each neuron generating a single spike during the forward process. 
% \PS{This "In particular" seems a bit unconnected to the previous sentence which talks about the concept of TTFS} 
A line of work focuses on training temporal-coded SNNs with backpropagation \citep{mostafa2017supervised,comsa2020temporal,bohte2000spikeprop,zhang2021rectified,xu2013supervised,shrestha2017robustness}, which highlights biological plausibility and efficiency of temporal coding. Much of the previous work has centered on developing improved synaptic models capable of effectively processing temporal information. For instance, \cite{mostafa2017supervised} employed non-leaky integrate-and-fire neurons to compute locally exact gradients for backpropagation, while \cite{comsa2020temporal} introduced the alpha-synaptic function to enhance SNNs' accuracy. Recently, \cite{zhang2021rectified} proposed a ReLU-like spike dynamics that effectively mitigates the dead neuron issue caused by the leaky nature of spike functions.

While advances in synaptic modeling have illuminated the understanding of neuronal dynamics, the exploration of network architecture in temporal SNNs has been relatively limited. In this paper, we explore architectural improvements of TTFS coding,  focusing on the role of skip connections in neural networks. Skip connections are a widely employed technique in ANNs, facilitating training and enhancing performance by allowing information to bypass certain layers. We examine two types of skip connection architectures: (1) addition-based skip connections, as proposed in ResNet \citep{he2016deep}, and (2) concatenation-based skip connections utilized in the ShuffleNetV2 architecture \citep{ma2018shufflenet}. We find that, when implemented in temporal SNN architectures, addition-based skip connections can introduce extra delays in the time dimension. Conversely, concatenation-based skip connections substantially reduce inference latency, but yield limited performance improvements due to discrepancies between the distributions from the convolution and skip branches. To augment the performance of concatenation-based skip connections, we propose a learnable delay for skip connections, which diminishes the distribution gap between the skip and convolutional branches, allowing for more effective information mixing between the two distributions.

In addition to our exploration of a new architecture for TTFS SNNs, we also investigate applications outside of image recognition, specifically in the realm of scientific machine-learning tasks.
We venture into the domain of time-reversal wave localization problems, a significant challenge in physics and engineering. This problem aims to trace back a wave's source given the wave shape at a later time \citep{kahana2022physically,bardos2002mathematical,givoli2012time}.
Through these experiments, we aim to demonstrate the versatility and potential of SNNs in various complex tasks, significantly expanding their applicability beyond traditional domains.

In summary, our contributions in this paper are threefold. 
(1) First, we explore the network architecture of temporal SNNs, with a particular emphasis on skip connections, examining both addition-based residual connections and concatenation-based skip connections in the context of temporal SNNs. 
(2) Second, we propose a learnable delay for skip connections to improve the performance of concatenation-based skip connections by reducing the distribution gap between skip and convolutional branches, enabling more effective information mixing. 
(3) Lastly, we extend the application of Time-To-First-Spike (TTFS) coding beyond image recognition to the time-reversal problem for source localization using wave signal. These contributions not only advance our understanding of network architecture in temporal SNNs but also broaden the potential applications of TTFS coding in various domains.

\section{Related Work}

\subsection{Spiking Neural Networks}
Spiking Neural Networks (SNNs), unlike traditional Artificial Neural Networks (ANNs), operate using temporal spikes, thereby offering a unique dimension of time in their operation \citep{roy2019towards,christensen20222022}. Among their components, the Leaky-Integrate-and-Fire (LIF) neuron is key as it functions as a non-linear activation unit. The LIF neurons stand out due to the ``memory" held within their membrane potential, where spikes are incrementally gathered. Once this potential exceeds a certain threshold, the neurons fire output spikes and the potential resets.

Training algorithms have been a primary focus of SNN research. Several methods proposed to address this involve transforming pre-trained ANNs into SNNs using weight or threshold balancing strategies \citep{sengupta2019going,han2020rmp,diehl2015fast,rueckauer2017conversion,li2021free}. While these techniques are generally effective, they require a multitude of timesteps to emulate float activations using binary spikes. A set of recent studies have suggested the use of surrogate functions to circumvent this non-differentiable backpropagation issue \citep{lee2016training,lee2020enabling,neftci2019surrogate,shrestha2018slayer,wu2018spatio,wu2021training,li2021differentiable,kim2022neural,wu2020progressive}. These methods, accounting for temporal dynamics during weight training, exhibit high performance and short latency.

Another significant aspect of SNN research pertains to the coding scheme. Several schemes have been proposed for image classification with SNNs. Burst coding, for example, communicates a burst of spikes within a short duration, enhancing synaptic communication reliability \citep{park2019fast}. Phase coding encodes temporal information into spike patterns based on a global oscillator \citep{montemurro2008phase}. Furthermore, rate coding has been applied to large-scale settings and is currently used by state-of-the-art methods \citep{diehl2015unsupervised,lee2016training,lee2020enabling}. This scheme generates a spike train over $T$ timesteps, where the total spike count reflects the magnitude of the input values. However, this generation of numerous spikes can pose issues for ultra-low power devices. To address this, Time-To-First-Spike (TTFS) coding has gained interest \citep{zhang2019tdsnn,mostafa2017supervised,comsa2020temporal}, as it generates a single spike per neuron, with spike latency inversely related to information importance. Despite progress in synaptic modeling, the architectural exploration of temporal SNNs remains limited. In this paper, we delve into the architectural enhancement of TTFS coding, specifically emphasizing the significance of skip connections in neural networks.

\vspace{4mm}
\subsection{Skip Connection Architecture}
The concept of skip connections or shortcut connections has been a cornerstone in the development of deep learning architectures, contributing significantly to the performance enhancement of various models.
\cite{he2016deep} proposed the ResNet architecture that employs skip connections to allow signals to bypass layers, directly flowing from one layer to a layer further into the network. This design helps to alleviate the problem of vanishing gradients in deep networks, thereby enabling the training of networks that are significantly deeper than those previously possible.
Furthermore, introduced by  \cite{srivastava2015training}, Highway Networks utilize gated skip connections, where the data flow is regulated by learned gating functions. This allows the network to learn to control the information flow dynamically.
Also, \citep{huang2017densely} proposed DenseNet, where each layer receives direct inputs from all preceding layers and passes down its own feature maps to all subsequent layers. This dense connectivity promotes feature reuse and substantially reduces the number of parameters.
In the ShuffleNetV2 architectures \citep{ma2018shufflenet}, channel shuffle operations and pointwise group convolutions are combined with skip connections to create highly efficient network architectures suitable for mobile devices.
The aforementioned architectures have shown the importance of skip connections in enhancing the performance of neural networks. However, most of these architectures have been designed for traditional Artificial Neural Networks (ANNs), and their application and efficacy in the context of Spiking Neural Networks (SNNs) remain to be thoroughly investigated.

\begin{figure*}[t]
\begin{center}
\def\arraystretch{0.5}
\begin{tabular}{@{\hskip 0.0\linewidth}c@{\hskip 0.0\linewidth}c@{\hskip 0.0\linewidth}c@{\hskip 0.0\linewidth}c@{}c}

\includegraphics[width=0.24\linewidth]{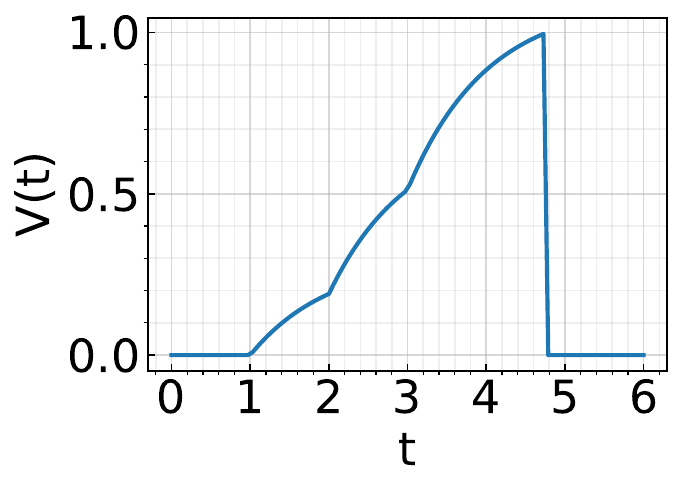} &
\includegraphics[width=0.245\linewidth]{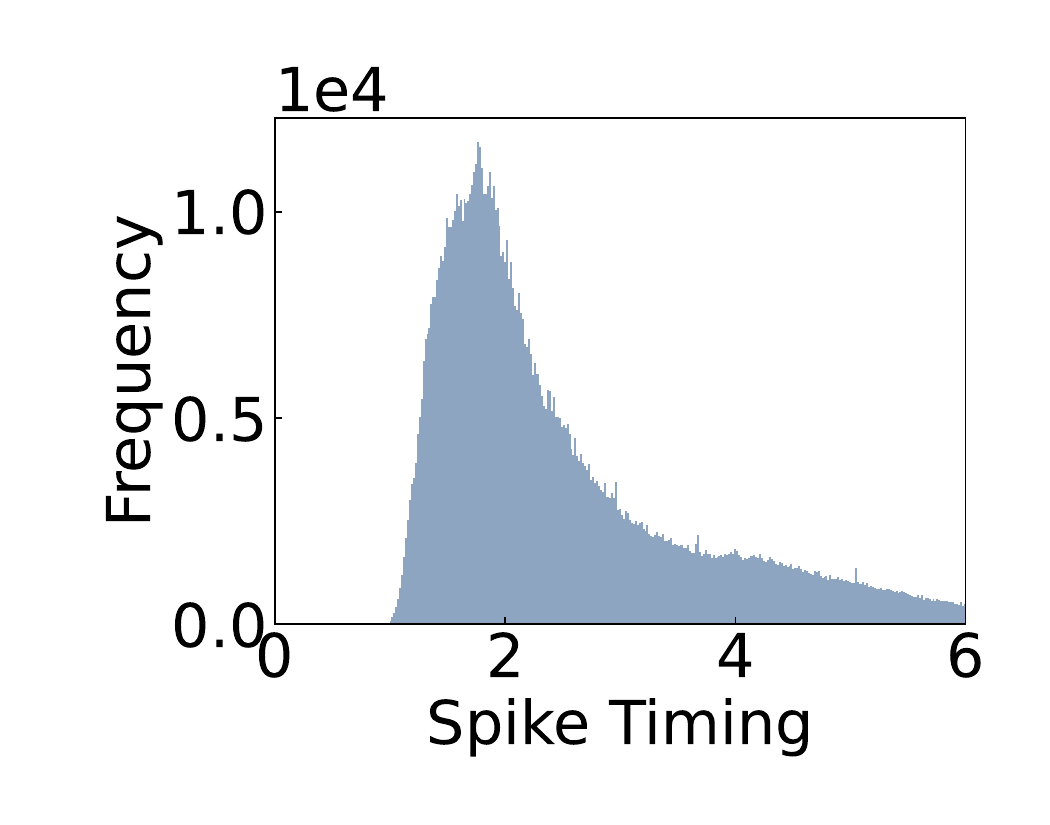}&
\includegraphics[width=0.245\linewidth]{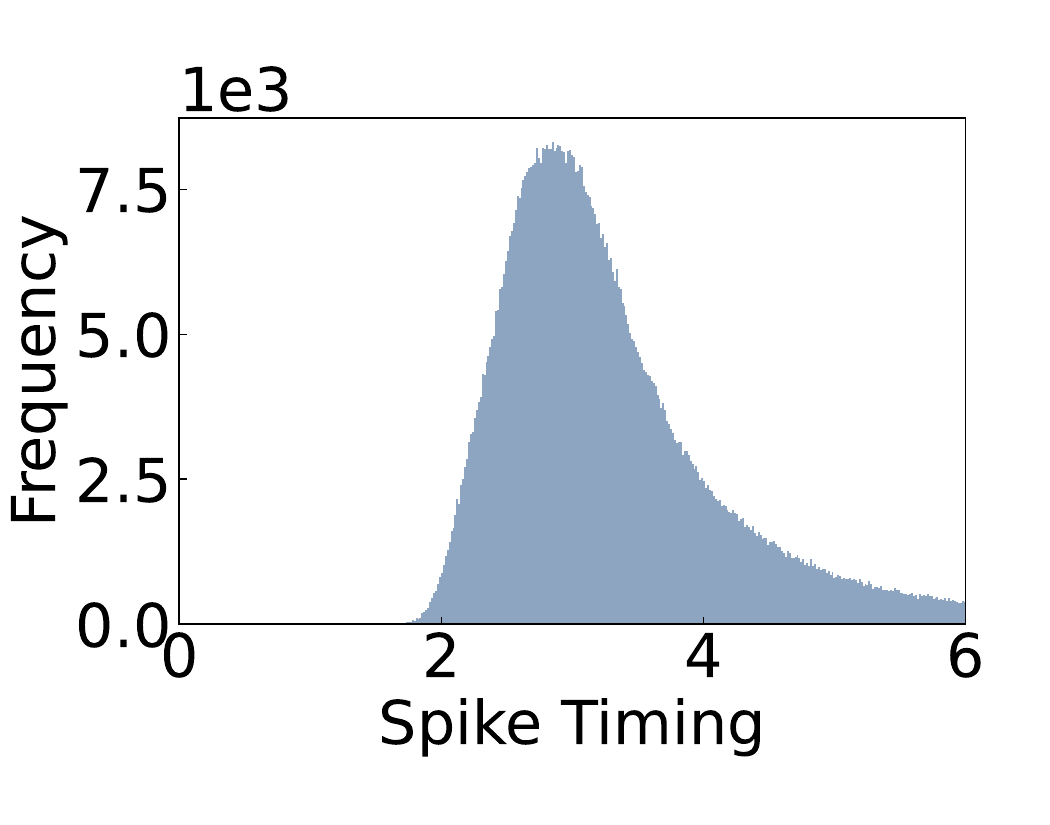}&
\includegraphics[width=0.245\linewidth]{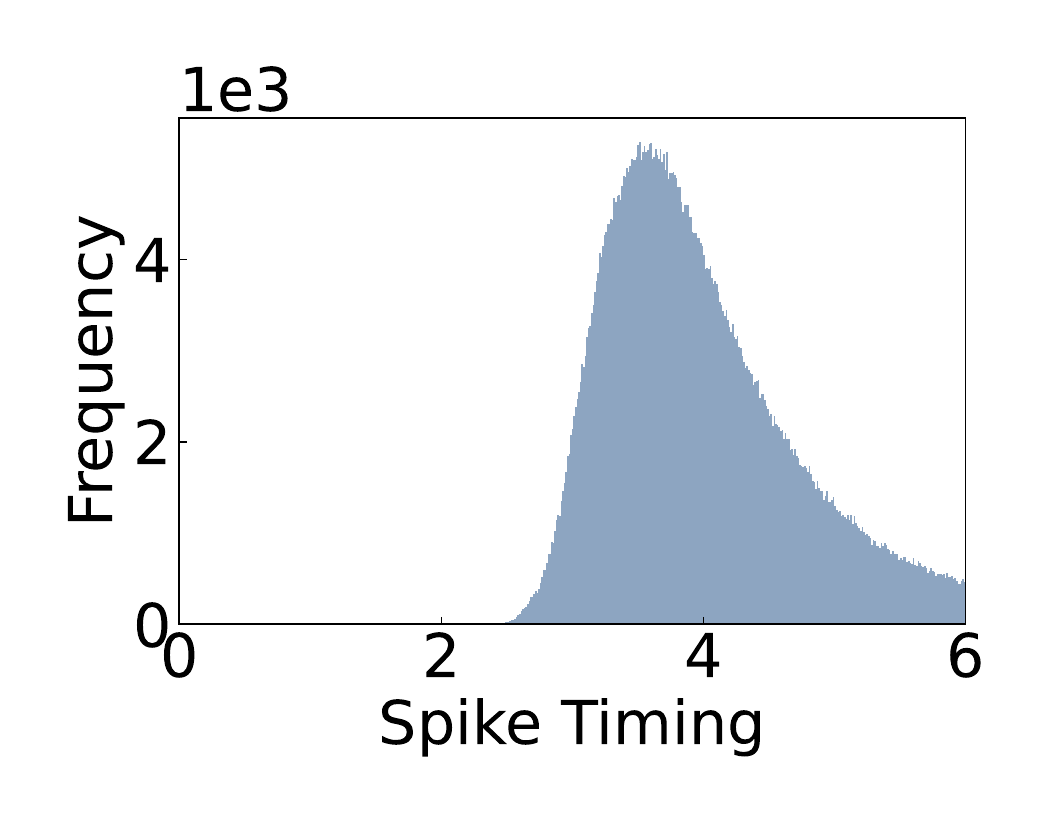}\\
\hspace{4mm} {(a)} & &{(b)}  & \\
\end{tabular}
% \vspace{-2mm}
\caption{ Illustration of a spike timing delay through layers. (a) Change of membrane potential when there are input spikes at times 1, 2, and 3. (b) We visualize the histogram of spike timing after Conv3, Conv4, and Conv5 layers.
% We use 5 Conv - 1 FC architectures.
}
 % \vspace{-1mm}
\label{fig:method:delay}
\end{center}
\end{figure*}

\section{Methodology}

\subsection{Temporal Neuron}
\label{sec:temporal_neuron}
Our neuron model is based on the non-leaky integrate-and-fire neurons proposed in \citep{mostafa2017supervised}.
The neuron employs exponentially decaying synaptic current kernels denoted as $\epsilon$. The influence of a spike occurring at time $t_k$ on the membrane potential can be expressed as
\begin{equation}
\frac{dV^j(t)}{dt} = \sum_i w_{ji} \sum_k \epsilon (t-t_k).
\label{eq:neuron_dynamic}
\end{equation}
Here, $V^j(t)$ represents the membrane potential of neuron $j$ at time $t$, and $w_{ji}$ is the weight connection between neuron $j$ and neuron $i$ in the preceding layer.
The synaptic current kernels, denoted by $\epsilon$, are defined as
\begin{equation}
\epsilon(x) = U(x)exp(-{x}), 
\label{eq:synaptic}
\end{equation}
where $U(x)$ is the step function. $U(x)=1$ when $x\ge 0$ and $U(x)=0$ otherwise, which ensures that the synaptic kernel accounts for the time subsequent to the input spike.
By considering both Eq. \ref{eq:neuron_dynamic} and Eq. \ref{eq:synaptic}, we can derive the equation of the membrane potential.
%\
We assume one neuron generates at most one spike, so the membrane potential $V(t)$ increases until there is an output spike. Then, we can write the membrane potential with $N$ input spikes 
\begin{equation}
    {V(t)} = \sum_{k=1}^N U(t-t_k) w_{k} (1-exp(-({t-t_k}))).
    \label{eq:membrane_dynamic}
\end{equation}
Here, $t_k$ is the spike timing of $k$-th spike and $w_k$ is the corresponding weight.

The neuron generates an output spike whenever the membrane potential has a higher value than $1$, thus, $V(t_{out}) \ge 1$, where $t_{out}$ means the output spike timing.   For all spike $t_k < t_{out}$, we refer to the set of input spike index as a casual set $C$, following the definition from the previous work \citep{mostafa2017supervised}.
\begin{equation}
    1 = \sum_{k\in C} w_{k} (1-exp(-({t_{out}-t_k}))).
    \label{eq:threshold}
\end{equation}
Then, we reorganize Eq. \ref{eq:threshold} for the spike timing.
\begin{equation}
    exp(t_{out}) = \frac{\sum_{k\in C} w_k exp(t_{k}) }{{\sum_{k\in C} w_k -1}}.
    \label{eq:output_eq}
\end{equation}
For simplicity, we transform $exp(t_i) \rightarrow z_i$, \ie, $z$-transformation \cite{weisstein2002z}.
Then, Eq. \ref{eq:output_eq} can be rewritten as:
\begin{equation}
    z_{out} = \frac{\sum_{k\in C} w_k z_{k} }{{\sum_{k\in C} w_k -1}}.
    \label{eq:z_output_eq}
\end{equation}
With TTFS coding, the networks determine the class of an image based on the neuron in the final layer that fires the earliest spike. For example, consider a scenario involving a 10-class classification problem. If the neuron corresponding to the ``dog" category emits the earliest spike in the final layer, the network immediately classifies the given image as a ``dog". 
This allows neural networks to have faster predictions, as the classification is determined as soon as the first spike is generated, without the need to wait for spikes from other neurons.

\begin{figure*}[t]
\begin{center}
\def\arraystretch{0.5}
\begin{tabular}{@{\hskip 0.0\linewidth}c@{\hskip 0.0\linewidth}c@{\hskip 0.0\linewidth}c@{\hskip 0.0\linewidth}c@{}c}

% \includegraphics[width=0.23\linewidth]{figures/hist_vis_skip2_res.pdf} &
% \includegraphics[width=0.23\linewidth]{figures/hist_vis_skip2_res.pdf} &
% \includegraphics[width=0.23\linewidth]{figures/hist_vis_conv2_res.pdf} &
% \includegraphics[width=0.23\linewidth]{figures/hist_vis_combine2_res.pdf}\\
%  &1 & 2& 3 &\\
 \includegraphics[width=0.99\linewidth]{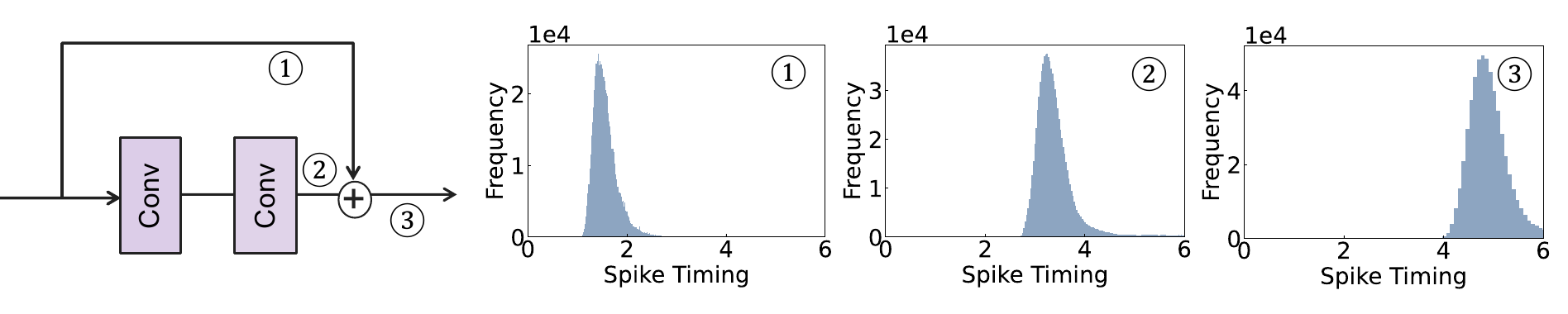} \\ 
{(a) Addition-based skip connection}   \\
\hspace{-3mm}\includegraphics[width=1.00\linewidth]{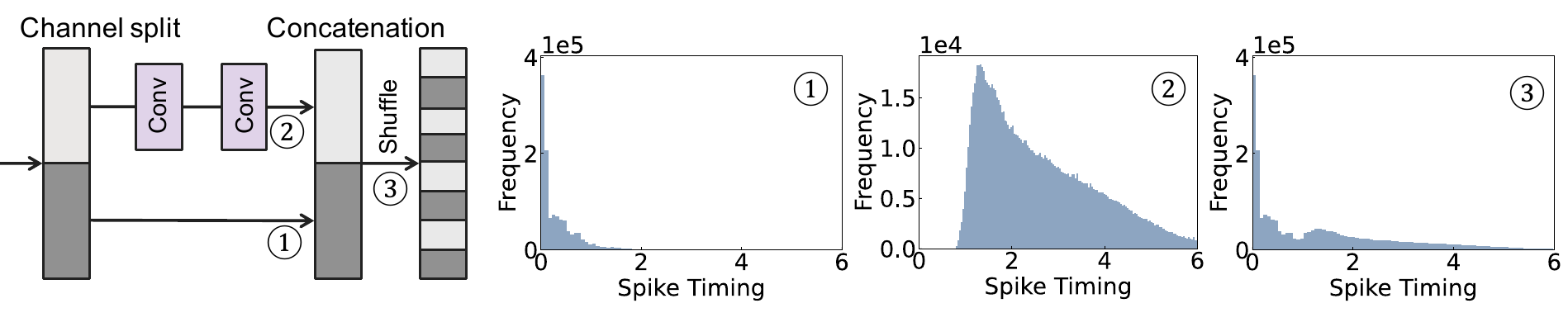} \\ 
{(b) Concatenation-based skip connection}   \\
\end{tabular}
% \vspace{-2mm}
\caption{ Illustration of a spike timing delay of two skip connection architectures.
We visualize the histogram of spike timing for \raisebox{.5pt}{\textcircled{\raisebox{-.9pt} {1}}}
 skip branch, \raisebox{.5pt}{\textcircled{\raisebox{-.9pt} {2}}} convolutional branch, and \raisebox{.5pt}{\textcircled{\raisebox{-.9pt} {3}}}  after combining the two branches. 
 % \PS{Are the three figures for the frequency of spike timing referring to Conv3, 4 and 5 layers like in Fig. 1?}
}
 \vspace{-3mm}
\label{fig:method:standard_res_shuffle}
\end{center}
\end{figure*}

\vspace{4mm}
\subsection{Observation: two types of skip connections}

\subsubsection{Temporal Delay in a Layer}
Our objective is to accelerate the first spike timing in the final layer, without compromising on the accuracy of the model.
In this context, one might wonder: \textit{what factors contribute to the delay in temporal coding?}
The delay is induced by spiking neurons where each neuron requires time to charge the membrane potential to generate the output spike, as demonstrated in Fig. \ref{fig:method:delay}(a).
Fig. \ref{fig:method:delay}(b) presents a histogram of spike timings, illustrating that the distribution shifts towards a later time as the layer goes deeper.
We aim to reduce this inherent temporal delay by bypassing the convolutional layers.

\vspace{4mm}
\subsubsection{Architectures}
\label{sssec:architecture}
We focus on the skip connection design in the temporal coding.  We examine two types of skip connection architectures: (1) addition-based skip connections, as seen in ResNet \citep{he2016deep}, and (2) concatenation-based skip connections utilized in the ShuffleNetV2 architecture \citep{ma2018shufflenet}. 
Addition-based skip connection architecture adds a skip connection to the main convolutional branch.
% \PS{Are the nonlinear branches always convolutional?}.
Let $X_l$ represent the input to $l$-th block, and $F(X_l)$ represent the non-linear transformation operations (\eg, convolution, batch normalization, 
 and non-linearity) within a residual block.
The operation of a residual block can be written as follows:
\begin{equation}
    X_{l+1} = F(X_l) + X_l.
    \label{eq:resnet}
\end{equation}
The overall operation of addition-based skip connection is illustrated in Fig. \ref{fig:method:standard_res_shuffle}(a). 
The major problem with this scheme is that adding two branches (\ie,  the skip connection and the convolutional branch) induces a significant delay in spike timing.
% \PS{Does the word "branches" here refer to the two convolutionalbranch or the two branches i.e., the skip connection and the nonlinear branch?} .
For example, adding a spike from a skip connection at time $t_A$ and a spike after convolutional operation at time $t_B$ results in the output at time $t_A+t_B$.
Therefore, the addition-based skip connection is not appropriate for TTFS coding.

On the other hand, a concatenation-based skip connection utilizes the channel split operation, where an input tensor is split into two parts along the channel dimension. One part is transformed through a series of operations while the other part passes a skip connection. These two parts are then concatenated and shuffled to ensure equal information sharing among channels.
Mathematically, for the input activation $X_l$ in $l$-th block, the operation of a concatenation-based skip connection can be represented as follows:
\begin{equation}
    X_{l} = [X_{l,1} ; X_{l,2}].
    \label{eq:shufflenet1}
\end{equation}
\begin{equation}
    X_{l+1} = \textup{Shuffle}(F(X_{l,1})|| X_{l,2}),
    \label{eq:shufflenet12}
\end{equation}
where $F(X_l)$ represents the non-linear transformation operation, and $X_{l,1}, X_{l,2}$ stand for two input tensors divided through channel dimension, respectively.
The overall illustration of the concatenation-based skip connection is shown in Fig. \ref{fig:method:standard_res_shuffle}(b).

Different from the addition-based skip connection, the concatenation-based skip connection allows spikes to pass directly through, thereby expediting the timing of spikes. However, this method does come with a notable disadvantage: a timing discrepancy between the spikes in the convolutional branch and those in the skip connection branch.
Specifically, we observe the distributions from the convolutional and skip connection branches have less overlap. 
This lack of overlap can make it difficult to integrate information effectively between the two distributions in the later layers. This is because of the inherent property of temporal neurons in TTFS coding. After a neuron spikes, there is no further activity, which implies a TTFS neuron will not consider any later input after it has output a spike. 
Consequently, these neurons tend to prioritize the earlier input, usually coming from the skip connection. This skewed consideration can potentially lead to a drop in accuracy, as vital information from later inputs could be overlooked. This observation underscores the importance of appropriately managing the timing of inputs in temporal SNNs to ensure effective information integration and high network performance.

To summarize, addition-based skip connections introduce additional timing delays in temporal SNNs. On the other hand, concatenation-based skip connections, despite speeding up the latency during inference, may overlook crucial information from the convolutional branch. 
% In our experiments section, we substantiate these observations through experiments conducted on various datasets, including those for image recognition and wave equation-solving tasks.

\begin{figure*}[t]
\begin{center}
\def\arraystretch{0.5}
\begin{tabular}{@{\hskip 0.0\linewidth}c@{\hskip 0.035\linewidth}c@{\hskip 0.015\linewidth}c@{\hskip 0.015\linewidth}c@{}c}
\includegraphics[width=0.95\linewidth]{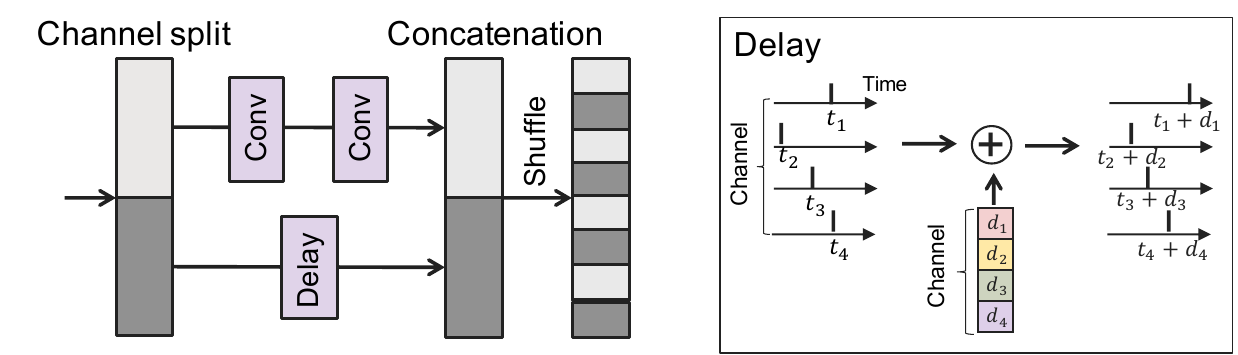} \\ 
\end{tabular}
% \vspace{-2mm}
\caption{ Illustration of the concatenation-based skip connection architecture with a delay block. 
}
 \vspace{-1mm}
\label{fig:method:delay_block}
\end{center}
\end{figure*}

\vspace{6mm}
\subsection{Adding Learnable Delay with a Skip Connection}

Hence, a question naturally arises: \textit{how can we improve the accuracy while reducing latency in TTFS?}
We focus on the problem of concatenation-based skip connections, \ie, timing discrepancy between the convolutional branch and the skip branch.
To address this problem, we introduce a delay to the skip connection, which is designed to minimize the timing disparity between the two branches. Fig. \ref{fig:method:delay_block} provides an illustration of the delay implementation within the skip connection, where a delay is added across each channel. This introduces a slight adjustment to the original concatenation-based skip connection architecture.
Initially, we partition the feature map across the channel dimension as follows:
\begin{equation}
    X_{l} = [X_{l,1} ; X_{l,2}].
    \label{eq:shufflenet1_delay}
\end{equation}
Subsequently, we apply a convolution layer $F(\cdot)$ to $X_{l,1}$ and a delay block $D(\cdot)$ to $X_{l,2}$. Then we concatenate the outputs from those branches:
\begin{equation}
    X_{l+1} = \textup{Shuffle}(F(X_{l,1})|| D(X_{l,2}|\theta_{l}))
    \label{eq:shufflenet2_delay},
\end{equation}
where the delay block can be written as follows:
\begin{equation}
   D(X|\theta_l) = X + \theta_l.
\end{equation}
Here, $\theta_l \in \mathbb{R}^{D}$ represents the parameters within the delay block in $l$-th layer, where $D$ is the channel dimension. This delay block applies distinct delays across each channel. We train these parameters alongside the other weight parameters within the neural network.
To further align the distributions from the convolutional layer and the skip connection, we introduce an additional loss constraint during optimization:
\begin{equation}
   \mathcal{L}_{overlap} = \sum_l ||Mean(F(X_{l,1})) -  Mean(D(X_{l,2}|\theta_l))||^2_2.
   \label{eq:}
\end{equation}
This loss function encourages the two distributions to converge, enhancing the efficacy of the information mixing between the convolutional and skip branches.

\vspace{8mm}
\subsection{Overall Optimization}

For the given image $X$, we convert the float input to the spike timing.
Similar to rate coding \citep{wu2019direct,wu2018spatio,kim2022rate}, we add one convolutional layer at the first layer, \ie, \textit{Conv-BN-ReLU}.
Then we directly utilize the output of the ReLU layer to the spike timing ($t = ReLU(BN(Conv(X)))$) as the output of the ReLU layer is always higher than zero. 
% \PS{How is this conversion achieved?}
We pass through the multiple temporal neuron layers as described in \ref{sec:temporal_neuron}.
In the output layer, the class probability is computed from the spike timing. The objective is to train the network such that the neuron associated with the correct class is the first to fire among all neurons.
This can be done by applying the cross-entropy loss to the output spike timing $O\in \mathbb{R}^C$ of the last layer:
\begin{equation}
   \mathcal{L}_{ce} = \sum_c  - y_c ln \frac{exp(-O_c)}{\sum_i exp(-O_i)}.
\end{equation}
Here, $y_c$ is a one-hot encoding of a class index, and $O_c$ denotes the $c$-th index of output neurons. Multiplying -1 with the output neuron is for assigning higher probability weighting for the early spike.

With a cross-entropy loss for classification, following the prior research \citep{mostafa2017supervised}, we introduce an additional term to the cost function that penalizes the input weight vectors of neurons whose sum is below 1 (denominator of Eq. \ref{eq:z_output_eq}). 
\begin{equation}
   \mathcal{L}_{weight} = \sum_l \sum_i max(0, 1-\sum_j w_{ij,l}).
\end{equation}
Here, $i$ is the neuron index of layer $l$, and $j$ is the input neuron index from the previous layer to neuron $i$.
% Throughout the training process, this term encourages the total of the weights in each neuron's input weight vector to exceed 1. Consequently, this guarantees that a neuron will generate a spike if all its input neurons do the same.
Overall, the total loss function is defined as follows.
\begin{equation}
   \mathcal{L}_{total} = \mathcal{L}_{ce} + \lambda_1 \mathcal{L}_{weight} + \lambda_2 \mathcal{L}_{overlap},
\end{equation}
where $\lambda_1$ and $\lambda_2$ are the hyperparameters for a trade-off between the losses.

\section{Experiments}
\subsection{Implementation Details}
We evaluate our method on MNIST \citep{lecun1998mnist} and Fashion-MNIST \citep{xiao2017fashion}.
We train the model with 128 batch samples using Adam optimizer with weight decay 1e-3. 
The initial learning rate is set to 6e-4 and decayed with cosine learning rate scheduling~\citep{loshchilov2016sgdr}. 
We set the total number of epochs to 100.
We set $\lambda_1$ and $\lambda_2$ to 1 and 1e-6, respectively. We use PyTorch for implementation. 
% The gradient is calculated as described in \citep{mostafa2017supervised}. 
% Additionally, floating-point values from the first convolutional layer are mapped into the floating time range of $0 \sim 6$. \PS{This is related to a previous question, what is this mapping?}

% \begin{table}[t]
% \addtolength{\tabcolsep}{1.0pt}
% \centering
% \caption{Architecture details.}
% \label{table:mnist_exp}
% \resizebox{0.9\textwidth}{!}{%
% \begin{tabular}{lc}
% \toprule
% Method   & Architecture\\
% \midrule
% Baseline  &  Conv(3,32)-Maxpool(2)-Conv(32,32, stride 2)- \\ 
% & Conv(32,32)-Conv(32,64, stride 2)-Conv(64,64)-FC(10)  \\
% \midrule
% Addition-based skip connection   &  Conv(3,32)-Maxpool(2)-Conv(32,32, stride 2)- \\ 
% & Conv(32,32)-Conv(32,64, stride 2)-Conv(64,64)-FC(10)  \\
%   \midrule
% Concatenation-based skip connection    & 97.0 \\
%  Concatenation-based skip connection  + Delay Block   & 98.0 \\
% \bottomrule
% \end{tabular}%
% }
% % \vspace{-0.5cm}
% \end{table}

\begin{figure*}[t]
\begin{center}
\def\arraystretch{0.5}
\begin{tabular}{@{\hskip 0.0\linewidth}c@{\hskip 0.035\linewidth}c@{\hskip 0.015\linewidth}c@{\hskip 0.015\linewidth}c@{}c}

\includegraphics[width=0.55\linewidth]{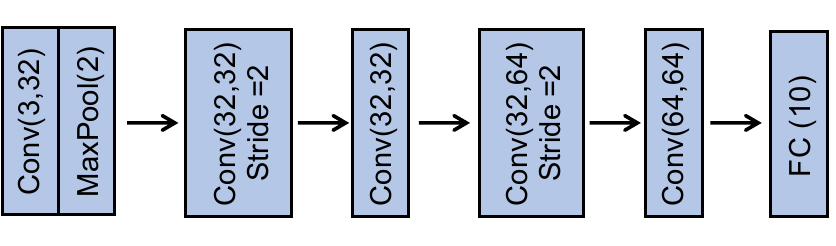} \\
{(a) Baseline CNN}\\\\
\includegraphics[width=0.59\linewidth]{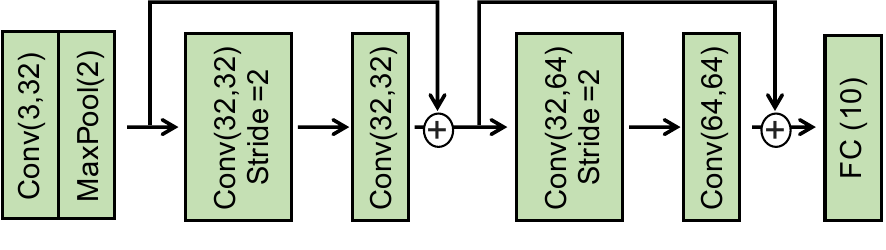} \\ 
{(b) Addition-based skip connection architecture}   \\
\includegraphics[width=0.69\linewidth]{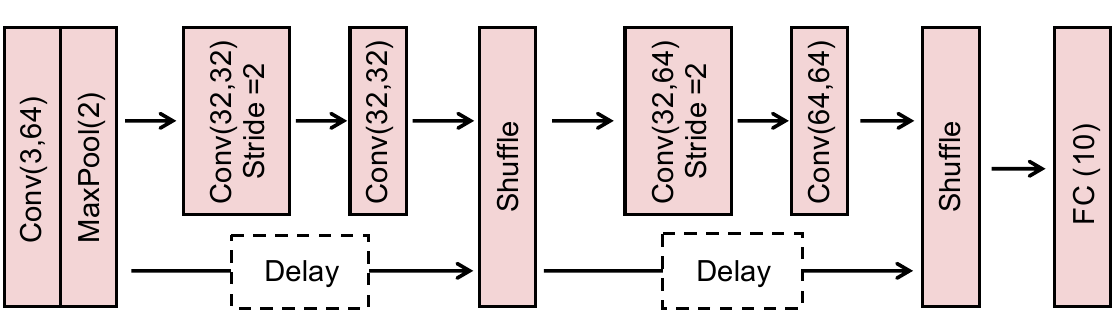} \\ 
{(c) Concatenation-based skip connection architecture}   \\
\end{tabular}
% \vspace{-2mm}
\caption{ Illustration of the architectures of Baseline CNN, Addition-based skip connection, and Concatenation-based skip connection.
}
 % \vspace{-1mm}
\label{fig:exp:architecture_detail}
\end{center}
\end{figure*}

\vspace{1mm}
\subsection{Experiments on Image Recognition}

Here, we compare the accuracy and latency across standard convolutional networks, addition-based skip connection architecture, and concatenation-based skip connection architecture.
More concretely, we construct baseline CNN architecture as follows:  \textit{Conv(3,32)-Maxpool(2)-Conv(32,32, stride 2)-Conv(32,32)-Conv(32,64, stride 2)-Conv(64,64)-FC(10)}.
For addition-based skip connection architecture, we add a residual connection for \textit{Conv(32,32, stride 2)-Conv(32,32)} block and \textit{Conv(32,64, stride 2)-Conv(64,64)}. In the skip connection, we apply a pooling layer to align the resolution between the skip connection and the convolutional branch. 
For a concatenation-based skip connection architecture, we also use a similar baseline CNN architecture.
% \PS{Does backbone mean baseline?}
Fig. \ref{fig:exp:architecture_detail} shows the details of the architectures. 
% Note that all architectures use temporal neurons proposed in \citep{mostafa2017supervised}.

To compare these architectures, we introduce latency as an additional metric, along with accuracy. In this context, latency is defined as the average time taken for the first spike to occur in the final layer. 
This is a particularly relevant measure for TTFS coding, as operations can be terminated as soon as the first spike occurs in the last layer. As such, we present both latency and accuracy in our results, offering a comprehensive understanding of the trade-off between speed and precision in these varying architectural designs.

In Table \ref{table:mnist_exp} and Table \ref{table:fmnist_exp}, we present the accuracy and latency results for the MNIST and Fashion-MNIST datasets respectively. Several key observations can be made from these results:
(1) The addition of skip connections significantly improves model accuracy compared to the baseline model. Specifically, the skip connection model improves performance by approximately 4\% for both the MNIST and Fashion-MNIST datasets.
(2) While the addition-based skip connection enhances accuracy, it also results in increased latency. This additional latency stems from the additive operation between the skip and convolutional branches, as discussed in Section \ref{sssec:architecture}.
(3) The concatenation-based skip connection model, without a delay block, achieves a reduction in latency. However, its performance is comparatively lower than that of the addition-based skip connection architecture.
(4) Incorporating a delay block into the concatenation-based skip connection model leads to improved performance. Notably, the addition of the delay block does not significantly increase latency for either dataset.
In summary, among the tested architectures, the concatenation-based skip connection model with a delay block provides the optimal balance of high performance and low latency.

\begin{table}[t]
\addtolength{\tabcolsep}{1.0pt}
\centering
\resizebox{0.9\textwidth}{!}
{
\begin{tabular}{lccc}
\toprule
Method   & Accuracy (\%) & Latency\\
\midrule
Baseline  (Convolution layers without skip connection) &   97.9 & 3.53 \\
Addition-based skip connection    & 98.4 & 4.57\\
Concatenation-based skip connection    & 98.4 & 2.16 \\
 Concatenation-based skip connection  + Delay Block   & 98.5 & 1.88\\
\bottomrule
\end{tabular}
}
\caption{Classification Accuracy (\%) and latency of skip connection architectures on the MNIST dataset.}
\label{table:mnist_exp}
% \vspace{-0.5cm}
\end{table}

\begin{table}[t]
\addtolength{\tabcolsep}{1.0pt}
\centering
\label{table:fmnist_exp}
\resizebox{0.9\textwidth}{!}{%
\begin{tabular}{lccc}
\toprule
Method   & Accuracy (\%) & Latency\\
\midrule
Baseline  (Convolution layers without skip connection) &   90.7 & 3.71 \\
Addition-based skip connection    & 91.1 & 4.89\\
Concatenation-based skip connection    & 90.6 & 2.28 \\
 Concatenation-based skip connection  + Delay Block   & 91.4 & 2.29\\
\bottomrule
\end{tabular}%
}
\caption{Classification Accuracy (\%) and latency of skip connection architectures on the Fashion-MNIST dataset.}
% \vspace{-0.5cm}
\end{table}

\vspace{4mm}
\subsection{Comparison with Previous Methods}

To establish the effectiveness of our proposed architecture, we draw comparisons between our model, Concat-based Skip Connection with Delay, and previous models. We focus on models that have been applied to the MNIST and Fashion-MNIST datasets.
Table \ref{table:dvs_acc} shows results for the MNIST dataset. Our model achieves an accuracy of 98.5\%, which is competitive with prior work.
For the Fashion-MNIST dataset, Table \ref{table:smnist_acc} illustrates the superior performance of our model, which achieves an accuracy of 91.4\%. The model by \cite{zhang2021rectified} is the next best performer with an accuracy of 90.1\%.
In both cases, our Concat-based Skip Connection with Delay architecture outperforms previous models, indicating its effectiveness in enhancing the performance of Spiking Neural Networks. This is particularly noteworthy for tasks that involve Time-To-First-Spike (TTFS) coding.
The success of the Concat-based Skip Connection with Delay model may be attributed to its ability to address the timing discrepancy between convolutional and skip branches, thus maintaining high performance while achieving low latency.
% \PS{Why are the accuracy results of our approach different in Tables 3 and 4 compared to Tables 1 and 2?}

\begin{table}[t]
\addtolength{\tabcolsep}{1.0pt}
\centering
\label{table:dvs_acc}
\resizebox{0.7\textwidth}{!}{%
\begin{tabular}{lcc}
\toprule
Method   & Coding & Accuracy (\%) \\
\midrule
Mostafa \etal \cite{mostafa2017supervised} & Temporal   & 97.5 \\
Comsa \etal  \cite{comsa2020temporal} & Temporal   & 97.9 \\
 Kheradpisheh \etal \cite{kheradpisheh2020temporal} & Temporal   & 97.4 \\
Kheradpisheh \etal     \cite{kheradpisheh2022bs4nn} & Temporal   & 97.0 \\
  Sakemi \etal   \cite{sakemi2021supervised} & Temporal  & 98.0 \\
    % \cite{zhang2021rectified} & Temporal   & 99.4 \\
    Ours (Concat-based Skip + Delay) & Temporal   & 98.5 \\
\bottomrule
\end{tabular}%
}
\caption{Accuracy (\%) comparison among the previous work on the MNIST dataset.}
% \vspace{-0.5cm}
\end{table}

\begin{table}[t]
\addtolength{\tabcolsep}{1.0pt}
\centering
\label{table:smnist_acc}
\resizebox{0.7\textwidth}{!}{%
\begin{tabular}{lcc}
\toprule
Method   & Coding & Accuracy (\%) \\
\midrule
   Hao \etal \cite{hao2020biologically} & Rate      & 85.3  \\
   Zhang \etal  \cite{zhang2020temporal} & Rate      & 89.5  \\
   Ranjan \etal  \cite{ranjan2020novel} & Rate   & 89.0 \\
Kheradpisheh \etal  \cite{kheradpisheh2020temporal} & Temporal   & 88.0 \\
   Kheradpisheh \etal  \cite{kheradpisheh2022bs4nn} & Temporal   & 87.3 \\
   Zhang \etal  \cite{zhang2021rectified} & Temporal   & 90.1 \\
    Ours (Concat-based Skip + Delay) & Temporal   & 91.4 \\
    % Wu \textit{et al.}   
\bottomrule
\end{tabular}%
}
\caption{
{
Accuracy (\%) comparison among the previous work on the Fashion-MNIST dataset.}}
% \vspace{-0.5cm}
\end{table}

\vspace{10mm}
\subsection{Experimental Analysis}

\begin{figure*}[t]
\begin{center}
\def\arraystretch{0.5}
\begin{tabular}{@{\hskip 0.13\linewidth}c@{\hskip 0.035\linewidth}c@{\hskip 0.015\linewidth}c@{\hskip 0.015\linewidth}c@{}c}

\includegraphics[width=0.83\linewidth]{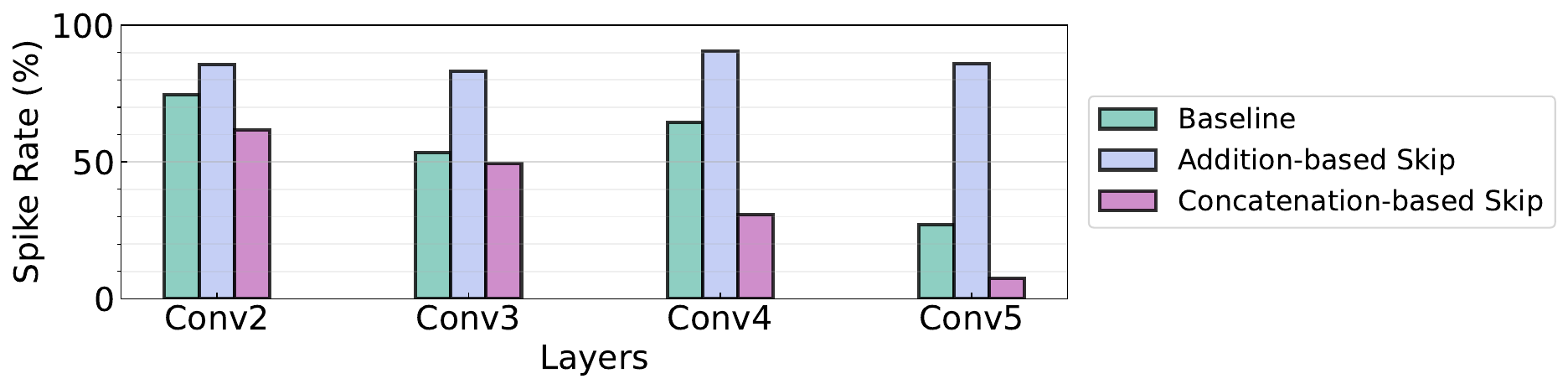} 
\end{tabular}
\vspace{-1mm}
\caption{ Spike rate in different architectures. We use the Fashion-MNIST dataset and measure the spike rate in Conv2 $\sim$ Conv5 layers.
}
 % \vspace{-1mm}
\label{fig:exp:spike_rate}
\end{center}
\end{figure*}

\vspace{0mm}
\subsubsection{Energy Efficiency: Spike Rate Comparison}
As highlighted in an earlier section, the ability to terminate operation within the networks as soon as the first spike at the last layer occurs is an inherent advantage of TTFS coding. To fully leverage this characteristic, we assess the number of spikes present in each layer when we implement an early exit strategy.
In Fig. \ref{fig:exp:spike_rate}, we visualize the spike rate of the baseline, addition-based skip connection architecture, and concatenation-based skip connection architecture. Here, we use the concatenation-based skip connection results with the added delay.
% \PS{I suppose that the concatenation-based skip connection results are with the added delay. If that is the case, we need to spell it out.} 
In this context, the spike rate refers to the proportion of firing neurons in a given layer. This metric is instrumental in understanding the speed and efficiency of each model, and consequently, its suitability for real-time or latency-sensitive applications.

The experimental results showcased in Fig. \ref{fig:exp:spike_rate} provide a clear comparison between the baseline model, the addition-based skip connection model, and the concatenation-based skip connection model in terms of spike rate at different layers of the network.
In the baseline model, the spike rates for conv2, conv3, conv4, and conv5 layers are 74.49\%, 53.45\%, 64.37\%, and 26.99\% respectively. For the addition-based skip connection model, the spike rates significantly increase, recorded at 85.5\%, 83.12\%, 90.61\%, and 86.02\% for conv2, conv3, conv4, and conv5 layers, respectively. These high spike rates demonstrate the model's heightened activity during the computation process.
On the other hand, the concatenation-based skip connection model exhibits considerably lower spike rates. Specifically, the rates for conv2, conv3, conv4, and conv5 layers are 61.59\%, 49.43\%, 30.67\%, and 7.59\% respectively. This indicates that fewer neurons are active during computation, thereby leading to lower latency and potentially faster computation times.

In summary, while the addition-based skip connection model tends to enhance activity across the network, the concatenation-based model successfully reduces the spike rate, potentially improving the efficiency of the network, particularly in latency-sensitive scenarios. These results further establish the importance of an appropriate skip connection strategy in designing efficient spiking neural networks.

\subsubsection{Comparison with ANN}
SNNs are renowned for their energy efficiency in comparison to traditional ANNs. To demonstrate this, we undertake a comparison of the approximate energy consumption between SNNs and ANNs, assuming that both are built using the same architecture.
Due to their event-driven nature and binary {1, 0} spike processing, SNNs are characterized by a reduced complexity of computations. More specifically, a multiply-accumulate (MAC) operation reduces in an SNN to a simple floating-point (FP) addition, thus necessitating only an accumulation (AC) operation. In contrast, traditional ANNs still require full MAC operations.
In line with previous studies such as \cite{park2020t2fsnn,lee2016training}, we estimate the energy consumption for SNNs by quantifying the total MAC operations involved. Using the standard 45nm CMOS technology as a reference point \cite{horowitz20141}, we assign the energy for MAC and AC operations as $E_{MAC}=4.6pJ$ and $E_{AC}=0.9pJ$ respectively. This shows that MAC operations consume approximately 5.11 times more energy than AC operations.
Since neurons in SNNs only consume energy whenever a neuron spikes, we multiply the spiking rate $R_s(l)$ at layer $l$  with FLOPs to obtain the SNN FLOPs count. 
The total inference energy of ANNs ($E_{ANN}$) and SNNs ($E_{SNN}$) are calculated by: 
$ E_{ANN} = \sum_{l} FLOPs(l)  \times E_{MAC}$ and 
$
E_{SNN} = \sum_{l} FLOPs(l) \times R_s(l) \times E_{AC}$, respectively. $FLOPs(l)$ represents the number of FLOPs in layer $l$.

The energy consumption comparison between our method and a conventional ANN is presented in Table \ref{table:exp:annsnn_energy}. Notably, while maintaining competitive performance, our method -- which combines concatenation-based skip connections with a delay block -- significantly reduces the relative energy cost. In contrast to the ANN's relative energy cost of 1, our method operates at just 13\% of the ANN's energy expenditure, thus demonstrating the superior energy efficiency of our approach.
% \PS{If an AC operation is 5 times more efficient than a MAC operation (20\% of energy compared to MAC), where does the extra 7\% energy saving come from for the final result of 13\%?}

\vspace{4mm}
\subsubsection{Delay Design}
There are several design variants for delay blocks. Here, we propose three types for comparison:
(1) Layer-wise delay applies the same delay across all neurons in one layer.
(2) Channel-wise delay adds a timing delay for each channel, which is used in our method.
(3) Pixel-wise delay adds a timing delay for each spatial location. For intermediate layers where the feature tensor size is $C \times H \times W$, we apply delay for each $H \times W$ location.   
% \PS{Is it clear how the pixel-wise delay is applied? I'm asking because the delay appears further in the SNN, while the pixels are in the input.}
Table \ref{table:delay_type} presents the comparative performance of these three delay block design variants, specifically evaluating their impact on accuracy and latency using the Fashion-MNIST dataset. The results suggest that the application of delay varies significantly in effect depending on its implementation level. Specifically, channel-wise delay, as employed in our method, demonstrated the highest accuracy and the lowest latency, indicating its effectiveness for the integration into concatenation-based skip connections. This demonstrates the potential benefits of applying unique delays to each channel, providing an effective balance between performance and computational efficiency.

\begin{table}[t]
\addtolength{\tabcolsep}{1.0pt}
\centering
\label{table:exp:annsnn_energy}
\resizebox{0.7\textwidth}{!}{%
\begin{tabular}{lcc}
\toprule
Method   & Relative Energy Cost & Accuracy (\%)  \\
\midrule
ANN & 1   & 92.27  \\
% ANN (iso-energy with concat-based SNN) & 0.13   & -\\
% SNN Baseline &  0.8  & 2.29 \\
Concat-based skip  + Delay  & 0.13   & 91.41 \\
\bottomrule
\end{tabular}%
}
\caption{Energy-efficiency comparison between ANN and SNN (with concatenation-based skip + delay) at inference. We use the Fashion-MNIST dataset for the experiments.}
% \vspace{-0.5cm}
\end{table}

\begin{table}[t]
\addtolength{\tabcolsep}{1.0pt}
\centering
\label{table:delay_type}
\resizebox{0.55\textwidth}{!}{%
\begin{tabular}{lcc}
\toprule
Method   & Accuracy (\%) & Latency \\
\midrule
Layer-wise Delay & 90.71   & 3.5\\
Channel-wise Delay & 91.41   & 2.29 \\
Pixel-wise Delay& 91.11   & 2.39 \\
\bottomrule
\end{tabular}%
}
\caption{Accuracy (\%) comparison among the different delay block designs. We use the Fashion-MNIST dataset for the experiments.}
% \vspace{-0.5cm}
\end{table}

\vspace{2mm}
\subsubsection{Analysis on Delay Value}

\begin{figure*}[t]
\begin{center}
\def\arraystretch{0.5}
\begin{tabular}{@{\hskip 0.05\linewidth}c@{\hskip 0.035\linewidth}c@{\hskip 0.015\linewidth}c@{\hskip 0.015\linewidth}c@{}c}

\includegraphics[width=0.37\linewidth]{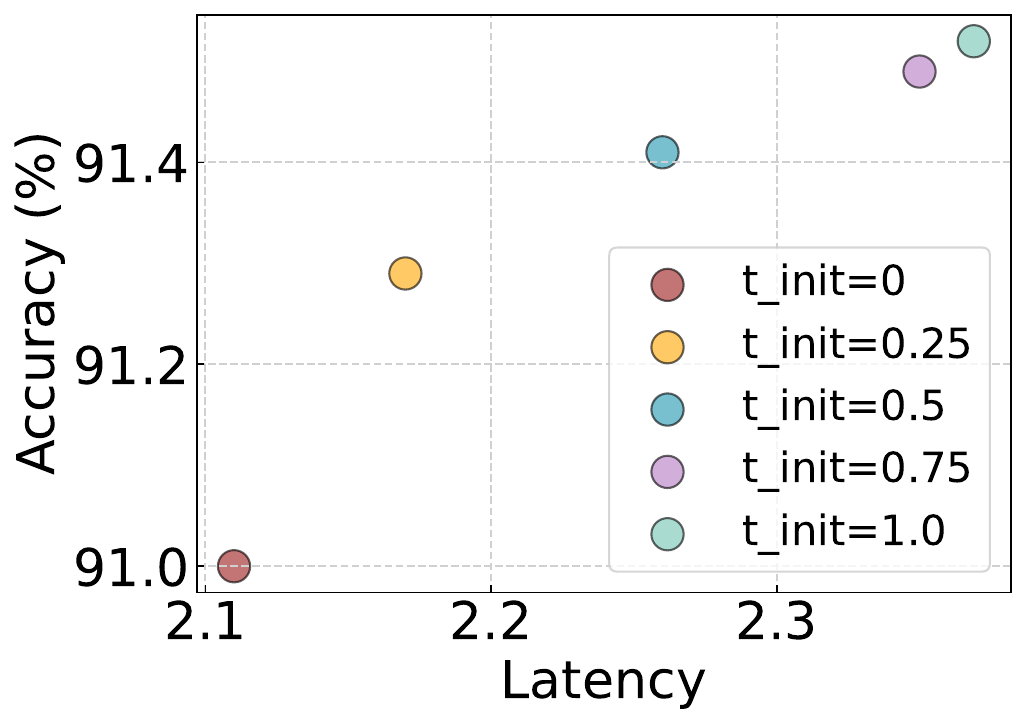}  & \hspace{10mm}\includegraphics[width=0.50\linewidth]{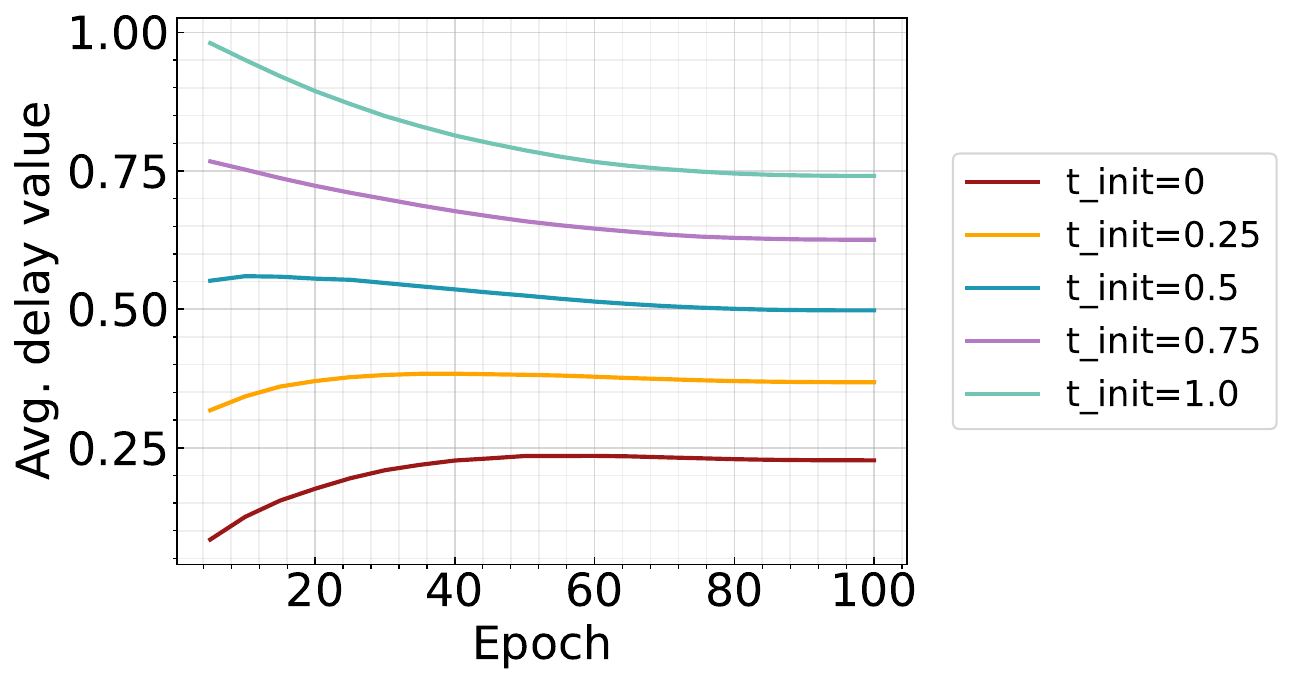} \\
% {(a) Training Loss}  &
% {(b) Training Accuracy}   \\
\end{tabular}
% \vspace{-2mm}
\caption{ \textbf{Left:} Accuracy and latency trade-off with respect to the time initialization in delay block. \textbf{Right:} Change of delay as training goes on. We train and test the model on Fashion-MNIST.
}
 % \vspace{-1mm}
\label{fig:exp:t_init}
\end{center}
\end{figure*}

In this section, we investigate the impact of different initial values within the delay block. For this purpose, we set the initial delay values to [0, 0.25, 0.5, 0.75, 1.0] and train the model accordingly.
As depicted in Fig. \ref{fig:exp:t_init} \textbf{Left}, we present the performance corresponding to each initialization time. It is observed that when initialized with a small delay, the resulting latency remains small. Conversely, when the delay value is initialized to be higher, the resultant latency increases, but this also leads to an enhancement in accuracy.
Fig. \ref{fig:exp:t_init} \textbf{Right} visualizes how these delay values fluctuate across epochs. We note that regardless of the initial value, all cases tend to gravitate towards a middle value over time. This results in high initial values decreasing over time, while lower initial values witness an increase.
% \PS{Can we reach a more concrete conclusion from these figures? We don't have to, just asking.}

\vspace{2mm}
\subsubsection{Advantage of Skip Connection: Training Stability}
\begin{figure*}[t]
\begin{center}
\def\arraystretch{0.5}
\begin{tabular}{@{\hskip -0.02\linewidth}c@{\hskip 0.09\linewidth}c@{\hskip 0.015\linewidth}c@{\hskip 0.015\linewidth}c@{}c}

\includegraphics[width=0.37\linewidth]{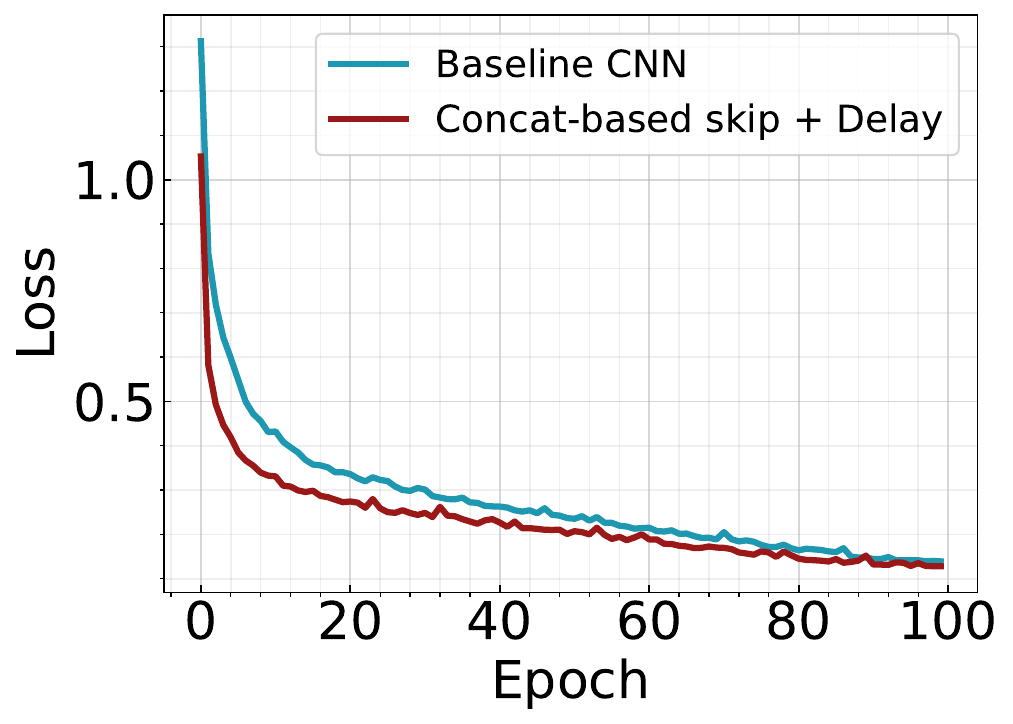}  & \includegraphics[width=0.37\linewidth]{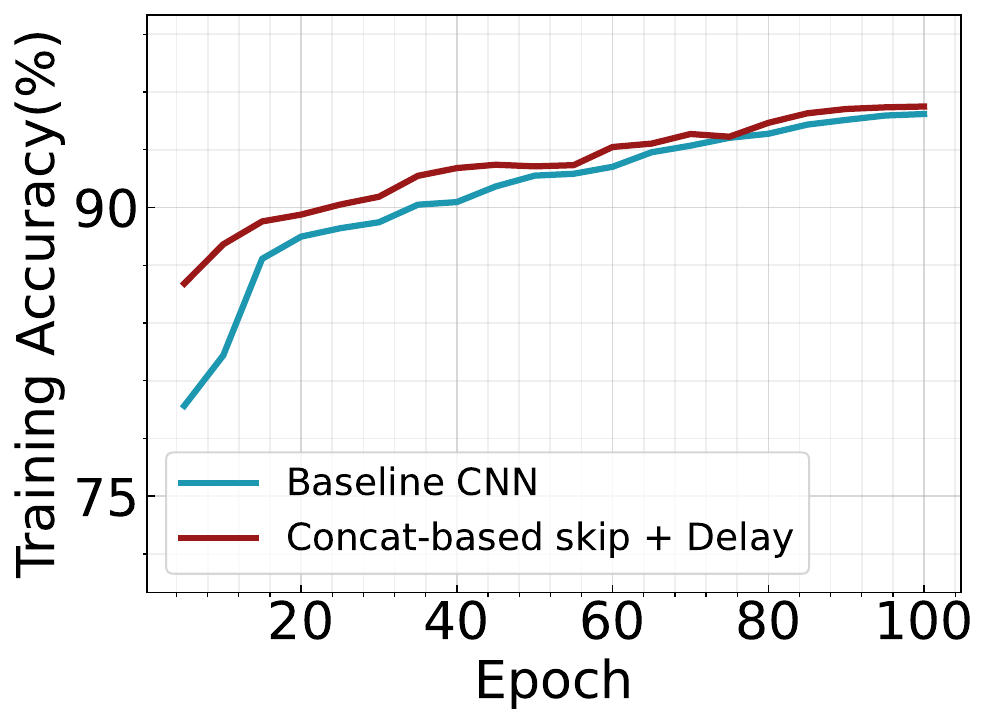} \\
{(a) Training Loss}  &
{(b) Training Accuracy}   \\
\end{tabular}
% \vspace{-2mm}
\caption{ Comparison of the (a) training loss
and (b) training accuracy across different architectures on Fashion-MNIST.
}
 % \vspace{-1mm}
\label{fig:exp:losacc_change}
\end{center}
\end{figure*}

Our analysis further explores the benefits of incorporating skip connections within the temporal SNN architecture. Prior research in the field of artificial neural networks (ANNs) suggests that the inclusion of skip connections enhances training stability and accelerates convergence speed \citep{he2016deep}. We scrutinize this premise in our current context by tracking and visualizing the evolution of training loss and accuracy over successive epochs, as depicted in Fig. \ref{fig:exp:losacc_change}. Our findings corroborate the aforementioned assertions, demonstrating that our optimized SNN architectures foster swift convergence while maintaining high test accuracy. This reinforces the value of skip connections as a significant contribution to the performance and efficiency of temporal SNNs.

\begin{figure*}[t]
\begin{center}
\def\arraystretch{0.5}
\begin{tabular}{@{\hskip 0.0\linewidth}c@{\hskip 0.035\linewidth}c@{\hskip 0.015\linewidth}c@{\hskip 0.015\linewidth}c@{}c}

\includegraphics[width=0.70\linewidth]{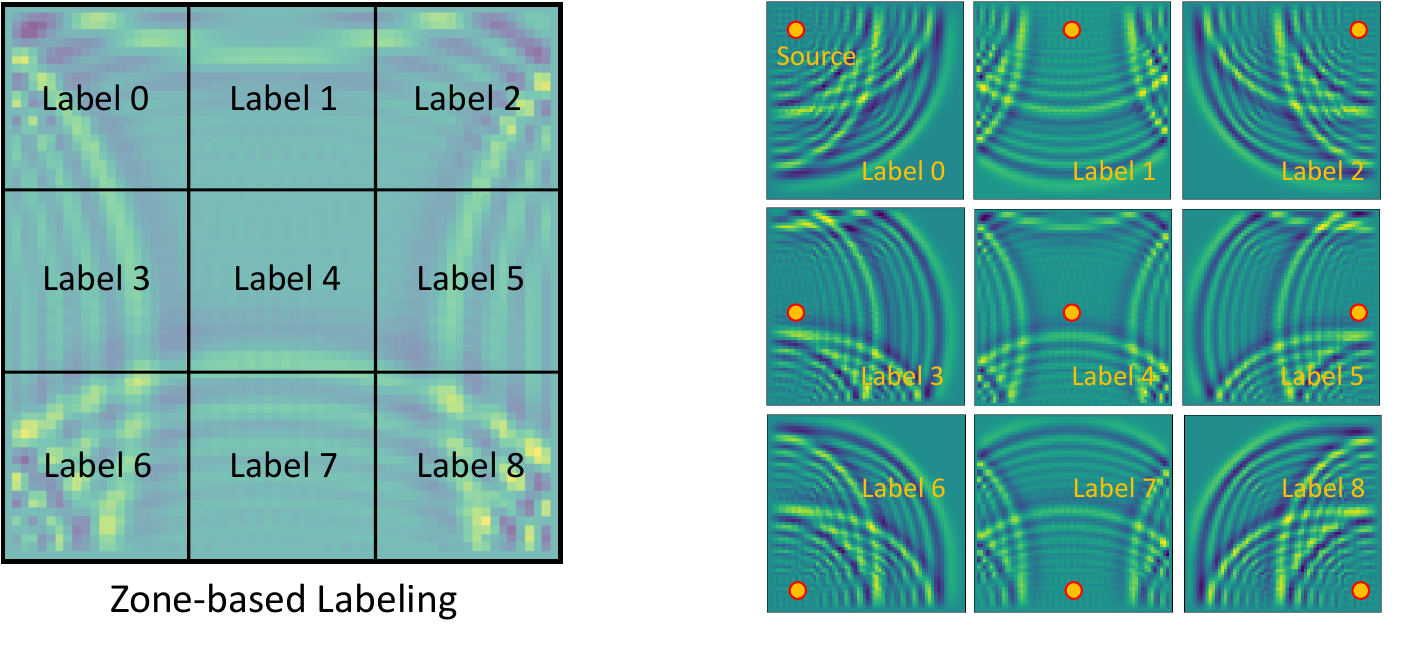} 
\end{tabular}
% \vspace{-2mm}
\caption{ The figure on the \textbf{left} illustrates an example of zone-based labeling for the source localization problem for the wave equation. The image is divided into $3 \times 3$ zones, with each zone assigned a distinct label. On the \textbf{right}, nine different wave images corresponding to each of the nine labels are displayed. Each image represents the wave shape at the 100-th time step, originating from a wave source located at the center of the respective zone.
}
 % \vspace{-1mm}
\label{fig:exp:wave_dataset}
\end{center}
\end{figure*}

\vspace{2mm}
\subsection{Experiments on Wave Equation}

% In broadening the scope of our investigation, we extend the application of TTFS coding to tasks within the domain of scientific machine learning. In particular, we explore the Time-Reversal procedure \citep{kahana2022physically,bardos2002mathematical,givoli2012time}, a technique that solves the wave equation in reverse temporal order to analyze past events. This technique leverages the reversibility property of the acoustic wave equation in a non-dissipative medium. Our objective is to evaluate the potential of TTFS SNNs as an approximation function for solving wave equations, thereby extending the utility of these networks beyond traditional image recognition tasks.

% The idea behind the TR procedure is to solve the wave equation “backwards in time” and examine events that occurred in
% the past, exploiting the reversibility property of the acoustic wave equation in a non-dissipative medium. The TR procedure
% is known to be robust to measurement noise [7,9,12,15]. We are interested in utilizing the TR process to compensate for
% the sensitivity to measurement noise, often experienced when using deep-learning algorithms

% \textbf{Should we have to write a wave equation here??}

In broadening the scope of our investigation, we extend the application of TTFS coding to tasks within the domain of scientific machine learning (SciML). An important topic in this field is how to approximate accurately solutions of partial differential equations. In particular, we are interested in solving the inverse (time-reversal) problem of locating sources in an underwater acoustic domain from measurements at a later time.
%such as somewhere in the Mediterranean sea. 
There has been a lot of research in this domain, with or without machine learning, as shown in this survey \citep{Grumiaux_2022}. The recent developments of Transformer-based architectures have recently been used for this task as well \citep{ovadia2023vito}. Another proposed method refers to using the Time-Reversal method incorporated with Machine learning based inference system \citep{kahana2022physically,bardos2002mathematical,givoli2012time}. Most methods still rely on ANNs that can be expensive to train.

The challenge herein is formulating this problem as a classification problem which aligns with the current implementation of the TTFS. The mathematical formulation of the wave problem we will investigate in this work is given by:

\begin{equation} \label{waveEq}
    \begin{cases}
        \ddot{u}(x, y, t) = c^2 \Delta u(x, y, t) & (x, y) \in \Omega = [x_{min}, x_{max}]\times [y_{min}, y_{max}], \quad t\in (0,T], \\
        u(x, y, 0) = u_0(x, y) & (x, y) \in \Omega, \\
        \dot{u}(x, y, 0) \equiv 0 & (x, y) \in \Omega, \\
        u(x, y, t) = 0 & (x, y) \in \partial \Omega,\quad t\in [0,T], \\
    \end{cases}
\end{equation}
where $u(x, y, t)$ is the acoustic wave pressure, and $c$ is the wave propagation velocity (assumed constant in this work and equal to $1484\frac{m}{s}$). A single dot over $u$ denotes a first derivative with respect to time. A double dot denotes a second derivative. The function $u_0$ is the initial condition for the wave propagation, determined by the source. This initial condition is taken as a small Gaussian eruption ($f(x)=e^{-(\ \hspace{-1mm}\frac{x-\overbar{x}}{0.05})\ ^2}$, $\overbar{x} = \frac{x_{max} - x_{min}}{2}$) that mimics a localized source (point source that has been smoothed to avoid numerical artifacts).  The goal is, from measurements of the wave pressure at some time $t$, to recover the location of the small initial eruption. This location is defined as a grid point. To turn it into a classification problem, we split the grid into zones, and infer the zone where the source is located. The more zones we use, the more precise localization we can achieve. However, more zones create a harder classification task, with growing computational demands.

\textbf{Dataset Configuration}: We generate a synthetic dataset based on the wave equation for a domain of $N_x \times N_y$ locations. We use a finite-difference central differences numerical scheme, preserving up to second order accuracy. We choose the ratio between the spatial discretization and the temporal discretization to satisfy the Courant-Freidrichs-Lewy condition (so that the scheme is stable). To create the synthetic dataset, for each sample, we choose a location for the source, use the solver to march in time and compute the wave pressure across the domain at the $100$-th time step. Then we create a label based on the location of the source, so the pairs consisting of an image of the pressure at the $100$-th time step and its corresponding label (source) form the dataset.

In detail, we posit the existence of a wave source at all locations within the domain, excluding a 10 pixels border around the outer boundary, thus resulting in $(N_x-10) \times (N_y-10)$ data samples. As mentioned above, for each wave source location, we compute the wave pressure at the $100$-th time step. To create the labels, we segment the domain into $M \times M$ zones and assign each source location a label from $0 \sim (M^2-1)$, as shown in Fig. \ref{fig:exp:wave_dataset} for a $3 \times 3$ zone-based labeling. The labeling process is the basic quantization of the domain into smaller segments, and the assigning label for each source according to the region it belongs to. For the $3 \times 3$ labeling, we have nine zones, meaning a total of nine classes. To get a more precise localization one can use $6 \times 6$ zones (as shown later), thus having 36 classes for the classification mechanism. Finally, we partition these data samples randomly into training and testing sets at an 80:20 ratio, respectively.

In Table \ref{table:exp:waveresults}, we report the accuracy and latency of the wave equation problem. We use the architecture shown in Fig. \ref{fig:exp:architecture_detail}.
Note that $6 \times 6$ zone-based labeling is a more difficult task than $3 \times 3$ zone-based labeling as the model requires classifying a larger number of zones.
% \PS{Need the details of the SNN implementation} 
We make the following observations:
(1) The general performance trend aligns with what we observed for image recognition tasks. Concatenation-based skip connection architectures, especially when paired with a delay block, show superior performance in terms of balancing high accuracy with lower latency. This supports the effectiveness of our proposed architecture for not only static image datasets but also for SciML tasks such as solving time-reversal problems for wave equations.
(2) The table also reveals that as we change the labeling strategy from $3 \times 3$ zones to $6 \times 6$ zones, we see an increase in latency across all models. This is intuitively right as a higher number of zones (classes) involves more computations and hence results in a longer latency.
In addition to the increased latency, a larger zone number also results in slightly reduced accuracy for all methods. This could be due to the increased complexity of the task with more zones, potentially requiring a more sophisticated model or additional training to achieve similar levels of accuracy as those observed for the smaller number of zones.
In Table \ref{table:exp:wave_annsnn_energy}, we compare the energy efficiency of SNN with ANN. Similar to image classification, our SNN consumes only 14\% of ANN energy, while sacrificing $\le 1\%$ accuracy.

\begin{table}[t]
\addtolength{\tabcolsep}{1.0pt}
\centering
\label{table:exp:waveresults}
\resizebox{0.95\textwidth}{!}{%
\begin{tabular}{lcccc}
\toprule
Method  & Labeling Strategy & Accuracy (\%) & Latency\\
\midrule
Baseline  (Convolution layers without skip connection)& $3 \times 3$ &  99.48 & 4.13 \\
Addition-based skip connection& $3 \times 3$    & 99.50 & 4.94\\
Concatenation-based skip connection & $3 \times 3$   & 96.36 & 1.06\\
 Concatenation-based skip connection  + Delay Block & $3 \times 3$  & 99.74 & 2.83\\
 \midrule
Baseline  (Convolution layers without skip connection)& $6 \times 6$ &   97.93& 4.57 \\
Addition-based skip connection& $6 \times 6$    & 97.94 & 4.87\\
Concatenation-based skip connection & $6 \times 6$   & 97.68 & 2.42 \\
 Concatenation-based skip connection  + Delay Block & $6 \times 6$  & 98.19 & 3.45\\
%  \midrule
% Baseline  (Convolution layers without skip connection)& $9 \times9$ &   90.70 & 3.7 \\
% Addition-based skip connection& $9 \times 9$    & 91.00 & 4.89\\
% Concatenation-based skip connection & $9 \times 9$   & 90.69 & 2.28 \\
%  Concatenation-based skip connection  + Delay Block & $9 \times 9$  & 91.41 & 2.29\\
\bottomrule
\end{tabular}%
}
\caption{Classification Accuracy (\%) and latency on time-reversal source localization problem for the wave equation. \textit{Labeling Strategy} means we use $M \times M$  zones for assigning a label to each source location. Fig. \ref{fig:exp:wave_dataset} shows an example of $3 \times 3$ zone-based labeling.}
% \vspace{-0.5cm}
\end{table}

\begin{table}[t]
\addtolength{\tabcolsep}{1.0pt}
\centering
\label{table:exp:wave_annsnn_energy}
\resizebox{0.7\textwidth}{!}{%
\begin{tabular}{lcc}
\toprule
Method   & Relative Energy Cost & Accuracy (\%)  \\
\midrule
ANN & 1   & 98.92  \\
% ANN (iso-energy with concat-based SNN) & 0.13   & -\\
% SNN Baseline &  0.8  & 2.29 \\
Concat-based skip  + Delay  & 0.14   & 98.19 \\
\bottomrule
\end{tabular}%
}
\caption{Energy-efficiency comparison between ANN and SNN (with concatenation-based skip + delay) at inference on the wave equation problem. We use the $6 \times 6$ zone-based labeling for the experiments.}
% \vspace{-0.5cm}
\end{table}

\section{Conclusion}
In conclusion, this study has made significant strides in the exploration of TTFS coding and the optimization of skip connection architectures for improving the efficiency and accuracy of SNNs. We discovered that while addition-based skip connections introduce temporal delays, concatenation-based skip connections tend to miss crucial information from the non-linear operation branch. 
% \PS{Here as before we focus on convolutional branch, but we had stated that this branch does not have to be convolutional, should we allow a more general name for the branch?}
To address these challenges, we proposed a novel approach that introduces a learnable delay for skip connections, bridging the gap between the spike timing discrepancies of the convolution and skip branches. We demonstrated that this method not only accelerates the first spike's timing but also maintains accuracy, offering an effective solution for faster prediction in TTFS coding.
We also extended our exploration to SciML tasks, unveiling the potential of TTFS coding beyond image recognition applications. Our findings suggest that there is room for further research in optimizing the network architecture of temporal SNNs, and we hope that our work will inspire new approaches and applications in this exciting field.
In the future, we aim to further improve the effectiveness of our proposed method and explore its applicability to even larger and more complex tasks. We believe that the continuing evolution of SNN architectures will significantly contribute to the advancement of low-power, efficient, and highly accurate artificial intelligence systems.

\section*{Acknowledgement} 
This work was supported in part by CoCoSys, a JUMP2.0 center sponsored by DARPA and SRC, the National Science Foundation (CAREER Award, Grant \#2312366, Grant \#2318152), TII (Abu Dhabi), and the U.S. Department of Energy, Advanced Scientific Computing Research program, under the Scalable, Efficient and Accelerated Causal Reasoning Operators, Graphs and Spikes for Earth and Embedded Systems (SEA-CROGS) project (Project No. 80278). Pacific Northwest National Laboratory (PNNL) is a multi-program national laboratory operated for the U.S. Department of Energy (DOE) by Battelle Memorial Institute under Contract No. DE-AC05-76RL01830.

\bibliography{sn-bibliography}% common bib file


\begin{thebibliography}{55}
\ifx \bisbn   \undefined \def \bisbn  #1{ISBN #1}\fi
\ifx \binits  \undefined \def \binits#1{#1}\fi
\ifx \bauthor  \undefined \def \bauthor#1{#1}\fi
\ifx \batitle  \undefined \def \batitle#1{#1}\fi
\ifx \bjtitle  \undefined \def \bjtitle#1{#1}\fi
\ifx \bvolume  \undefined \def \bvolume#1{\textbf{#1}}\fi
\ifx \byear  \undefined \def \byear#1{#1}\fi
\ifx \bissue  \undefined \def \bissue#1{#1}\fi
\ifx \bfpage  \undefined \def \bfpage#1{#1}\fi
\ifx \blpage  \undefined \def \blpage #1{#1}\fi
\ifx \burl  \undefined \def \burl#1{\textsf{#1}}\fi
\ifx \doiurl  \undefined \def \doiurl#1{\url{https://doi.org/#1}}\fi
\ifx \betal  \undefined \def \betal{\textit{et al.}}\fi
\ifx \binstitute  \undefined \def \binstitute#1{#1}\fi
\ifx \binstitutionaled  \undefined \def \binstitutionaled#1{#1}\fi
\ifx \bctitle  \undefined \def \bctitle#1{#1}\fi
\ifx \beditor  \undefined \def \beditor#1{#1}\fi
\ifx \bpublisher  \undefined \def \bpublisher#1{#1}\fi
\ifx \bbtitle  \undefined \def \bbtitle#1{#1}\fi
\ifx \bedition  \undefined \def \bedition#1{#1}\fi
\ifx \bseriesno  \undefined \def \bseriesno#1{#1}\fi
\ifx \blocation  \undefined \def \blocation#1{#1}\fi
\ifx \bsertitle  \undefined \def \bsertitle#1{#1}\fi
\ifx \bsnm \undefined \def \bsnm#1{#1}\fi
\ifx \bsuffix \undefined \def \bsuffix#1{#1}\fi
\ifx \bparticle \undefined \def \bparticle#1{#1}\fi
\ifx \barticle \undefined \def \barticle#1{#1}\fi
\bibcommenthead
\ifx \bconfdate \undefined \def \bconfdate #1{#1}\fi
\ifx \botherref \undefined \def \botherref #1{#1}\fi
\ifx \url \undefined \def \url#1{\textsf{#1}}\fi
\ifx \bchapter \undefined \def \bchapter#1{#1}\fi
\ifx \bbook \undefined \def \bbook#1{#1}\fi
\ifx \bcomment \undefined \def \bcomment#1{#1}\fi
\ifx \oauthor \undefined \def \oauthor#1{#1}\fi
\ifx \citeauthoryear \undefined \def \citeauthoryear#1{#1}\fi
\ifx \endbibitem  \undefined \def \endbibitem {}\fi
\ifx \bconflocation  \undefined \def \bconflocation#1{#1}\fi
\ifx \arxivurl  \undefined \def \arxivurl#1{\textsf{#1}}\fi
\csname PreBibitemsHook\endcsname

%%% 1
\bibitem[\protect\citeauthoryear{Roy et~al.}{2019}]{roy2019towards}
\begin{barticle}
\bauthor{\bsnm{Roy}, \binits{K.}},
\bauthor{\bsnm{Jaiswal}, \binits{A.}},
\bauthor{\bsnm{Panda}, \binits{P.}}:
\batitle{Towards spike-based machine intelligence with neuromorphic computing}.
\bjtitle{Nature}
\bvolume{575}(\bissue{7784}),
\bfpage{607}--\blpage{617}
(\byear{2019})
\end{barticle}
\endbibitem

%%% 2
\bibitem[\protect\citeauthoryear{Christensen
  et~al.}{2022}]{christensen20222022}
\begin{botherref}
\oauthor{\bsnm{Christensen}, \binits{D.V.}},
\oauthor{\bsnm{Dittmann}, \binits{R.}},
\oauthor{\bsnm{Linares-Barranco}, \binits{B.}},
\oauthor{\bsnm{Sebastian}, \binits{A.}},
\oauthor{\bsnm{Le~Gallo}, \binits{M.}},
\oauthor{\bsnm{Redaelli}, \binits{A.}},
\oauthor{\bsnm{Slesazeck}, \binits{S.}},
\oauthor{\bsnm{Mikolajick}, \binits{T.}},
\oauthor{\bsnm{Spiga}, \binits{S.}},
\oauthor{\bsnm{Menzel}, \binits{S.}}, et al.:
2022 roadmap on neuromorphic computing and engineering.
Neuromorphic Computing and Engineering
(2022)
\end{botherref}
\endbibitem

%%% 3
\bibitem[\protect\citeauthoryear{Panda et~al.}{2020}]{panda2020toward}
\begin{botherref}
\oauthor{\bsnm{Panda}, \binits{P.}},
\oauthor{\bsnm{Aketi}, \binits{S.A.}},
\oauthor{\bsnm{Roy}, \binits{K.}}:
Toward scalable, efficient, and accurate deep spiking neural networks with
  backward residual connections, stochastic softmax, and hybridization.
Frontiers in Neuroscience
\textbf{14}
(2020)
\end{botherref}
\endbibitem

%%% 4
\bibitem[\protect\citeauthoryear{Cao et~al.}{2015}]{cao2015spiking}
\begin{barticle}
\bauthor{\bsnm{Cao}, \binits{Y.}},
\bauthor{\bsnm{Chen}, \binits{Y.}},
\bauthor{\bsnm{Khosla}, \binits{D.}}:
\batitle{Spiking deep convolutional neural networks for energy-efficient object
  recognition}.
\bjtitle{International Journal of Computer Vision}
\bvolume{113}(\bissue{1}),
\bfpage{54}--\blpage{66}
(\byear{2015})
\end{barticle}
\endbibitem

%%% 5
\bibitem[\protect\citeauthoryear{Diehl and Cook}{2015}]{diehl2015unsupervised}
\begin{barticle}
\bauthor{\bsnm{Diehl}, \binits{P.U.}},
\bauthor{\bsnm{Cook}, \binits{M.}}:
\batitle{Unsupervised learning of digit recognition using
  spike-timing-dependent plasticity}.
\bjtitle{Frontiers in computational neuroscience}
\bvolume{9},
\bfpage{99}
(\byear{2015})
\end{barticle}
\endbibitem

%%% 6
\bibitem[\protect\citeauthoryear{Comsa et~al.}{2020}]{comsa2020temporal}
\begin{bchapter}
\bauthor{\bsnm{Comsa}, \binits{I.M.}},
\bauthor{\bsnm{Fischbacher}, \binits{T.}},
\bauthor{\bsnm{Potempa}, \binits{K.}},
\bauthor{\bsnm{Gesmundo}, \binits{A.}},
\bauthor{\bsnm{Versari}, \binits{L.}},
\bauthor{\bsnm{Alakuijala}, \binits{J.}}:
\bctitle{Temporal coding in spiking neural networks with alpha synaptic
  function}.
In: \bbtitle{ICASSP 2020-2020 IEEE International Conference on Acoustics,
  Speech and Signal Processing (ICASSP)},
pp. \bfpage{8529}--\blpage{8533}
(\byear{2020}).
\bcomment{IEEE}
\end{bchapter}
\endbibitem

%%% 7
\bibitem[\protect\citeauthoryear{Park et~al.}{2019}]{park2019fast}
\begin{bchapter}
\bauthor{\bsnm{Park}, \binits{S.}}, \betal:
\bctitle{Fast and efficient information transmission with burst spikes in deep
  spiking neural networks}.
In: \bbtitle{2019 56th ACM/IEEE Design Automation Conference (DAC)},
pp. \bfpage{1}--\blpage{6}
(\byear{2019}).
\bcomment{IEEE}
\end{bchapter}
\endbibitem

%%% 8
\bibitem[\protect\citeauthoryear{Guo et~al.}{2021}]{guo2021neural}
\begin{barticle}
\bauthor{\bsnm{Guo}, \binits{W.}}, \betal:
\batitle{Neural coding in spiking neural networks: A comparative study for
  robust neuromorphic systems}.
\bjtitle{Frontiers in Neuroscience}
\bvolume{15},
\bfpage{212}
(\byear{2021})
\end{barticle}
\endbibitem

%%% 9
\bibitem[\protect\citeauthoryear{Wu et~al.}{2019}]{wu2019direct}
\begin{bchapter}
\bauthor{\bsnm{Wu}, \binits{Y.}},
\bauthor{\bsnm{Deng}, \binits{L.}},
\bauthor{\bsnm{Li}, \binits{G.}},
\bauthor{\bsnm{Zhu}, \binits{J.}},
\bauthor{\bsnm{Xie}, \binits{Y.}},
\bauthor{\bsnm{Shi}, \binits{L.}}:
\bctitle{Direct training for spiking neural networks: Faster, larger, better}.
In: \bbtitle{Proceedings of the AAAI Conference on Artificial Intelligence},
vol. \bseriesno{33},
pp. \bfpage{1311}--\blpage{1318}
(\byear{2019})
\end{bchapter}
\endbibitem

%%% 10
\bibitem[\protect\citeauthoryear{Zheng et~al.}{2020}]{zheng2020going}
\begin{botherref}
\oauthor{\bsnm{Zheng}, \binits{H.}},
\oauthor{\bsnm{Wu}, \binits{Y.}},
\oauthor{\bsnm{Deng}, \binits{L.}},
\oauthor{\bsnm{Hu}, \binits{Y.}},
\oauthor{\bsnm{Li}, \binits{G.}}:
Going deeper with directly-trained larger spiking neural networks.
arXiv preprint arXiv:2011.05280
(2020)
\end{botherref}
\endbibitem

%%% 11
\bibitem[\protect\citeauthoryear{Zhang and Li}{2020}]{zhang2020temporal}
\begin{botherref}
\oauthor{\bsnm{Zhang}, \binits{W.}},
\oauthor{\bsnm{Li}, \binits{P.}}:
Temporal spike sequence learning via backpropagation for deep spiking neural
  networks.
arXiv preprint arXiv:2002.10085
(2020)
\end{botherref}
\endbibitem

%%% 12
\bibitem[\protect\citeauthoryear{Fang et~al.}{2020}]{fang2020incorporating}
\begin{botherref}
\oauthor{\bsnm{Fang}, \binits{W.}}, et al.:
Incorporating learnable membrane time constant to enhance learning of spiking
  neural networks.
arXiv preprint arXiv:2007.05785
(2020)
\end{botherref}
\endbibitem

%%% 13
\bibitem[\protect\citeauthoryear{Lee et~al.}{2020}]{lee2020enabling}
\begin{botherref}
\oauthor{\bsnm{Lee}, \binits{C.}},
\oauthor{\bsnm{Sarwar}, \binits{S.S.}},
\oauthor{\bsnm{Panda}, \binits{P.}},
\oauthor{\bsnm{Srinivasan}, \binits{G.}},
\oauthor{\bsnm{Roy}, \binits{K.}}:
Enabling spike-based backpropagation for training deep neural network
  architectures.
Frontiers in Neuroscience
\textbf{14}
(2020)
\end{botherref}
\endbibitem

%%% 14
\bibitem[\protect\citeauthoryear{Montemurro et~al.}{2008}]{montemurro2008phase}
\begin{barticle}
\bauthor{\bsnm{Montemurro}, \binits{M.A.}},
\bauthor{\bsnm{Rasch}, \binits{M.J.}},
\bauthor{\bsnm{Murayama}, \binits{Y.}},
\bauthor{\bsnm{Logothetis}, \binits{N.K.}},
\bauthor{\bsnm{Panzeri}, \binits{S.}}:
\batitle{Phase-of-firing coding of natural visual stimuli in primary visual
  cortex}.
\bjtitle{Current biology}
\bvolume{18}(\bissue{5}),
\bfpage{375}--\blpage{380}
(\byear{2008})
\end{barticle}
\endbibitem

%%% 15
\bibitem[\protect\citeauthoryear{Zhang et~al.}{2019}]{zhang2019tdsnn}
\begin{bchapter}
\bauthor{\bsnm{Zhang}, \binits{L.}},
\bauthor{\bsnm{Zhou}, \binits{S.}},
\bauthor{\bsnm{Zhi}, \binits{T.}},
\bauthor{\bsnm{Du}, \binits{Z.}},
\bauthor{\bsnm{Chen}, \binits{Y.}}:
\bctitle{Tdsnn: From deep neural networks to deep spike neural networks with
  temporal-coding}.
In: \bbtitle{Proceedings of the AAAI Conference on Artificial Intelligence},
vol. \bseriesno{33},
pp. \bfpage{1319}--\blpage{1326}
(\byear{2019})
\end{bchapter}
\endbibitem

%%% 16
\bibitem[\protect\citeauthoryear{Rueckauer and
  Liu}{2018}]{rueckauer2018conversion}
\begin{bchapter}
\bauthor{\bsnm{Rueckauer}, \binits{B.}},
\bauthor{\bsnm{Liu}, \binits{S.-C.}}:
\bctitle{Conversion of analog to spiking neural networks using sparse temporal
  coding}.
In: \bbtitle{2018 IEEE International Symposium on Circuits and Systems
  (ISCAS)},
pp. \bfpage{1}--\blpage{5}
(\byear{2018}).
\bcomment{IEEE}
\end{bchapter}
\endbibitem

%%% 17
\bibitem[\protect\citeauthoryear{Mostafa}{2017}]{mostafa2017supervised}
\begin{barticle}
\bauthor{\bsnm{Mostafa}, \binits{H.}}:
\batitle{Supervised learning based on temporal coding in spiking neural
  networks}.
\bjtitle{IEEE transactions on neural networks and learning systems}
\bvolume{29}(\bissue{7}),
\bfpage{3227}--\blpage{3235}
(\byear{2017})
\end{barticle}
\endbibitem

%%% 18
\bibitem[\protect\citeauthoryear{Bohte et~al.}{2000}]{bohte2000spikeprop}
\begin{bchapter}
\bauthor{\bsnm{Bohte}, \binits{S.M.}},
\bauthor{\bsnm{Kok}, \binits{J.N.}},
\bauthor{\bsnm{La~Poutr{\'e}}, \binits{J.A.}}:
\bctitle{Spikeprop: backpropagation for networks of spiking neurons.}
In: \bbtitle{ESANN},
vol. \bseriesno{48},
pp. \bfpage{419}--\blpage{424}
(\byear{2000}).
\bcomment{Bruges}
\end{bchapter}
\endbibitem

%%% 19
\bibitem[\protect\citeauthoryear{Zhang et~al.}{2021}]{zhang2021rectified}
\begin{barticle}
\bauthor{\bsnm{Zhang}, \binits{M.}},
\bauthor{\bsnm{Wang}, \binits{J.}},
\bauthor{\bsnm{Wu}, \binits{J.}},
\bauthor{\bsnm{Belatreche}, \binits{A.}},
\bauthor{\bsnm{Amornpaisannon}, \binits{B.}},
\bauthor{\bsnm{Zhang}, \binits{Z.}},
\bauthor{\bsnm{Miriyala}, \binits{V.P.K.}},
\bauthor{\bsnm{Qu}, \binits{H.}},
\bauthor{\bsnm{Chua}, \binits{Y.}},
\bauthor{\bsnm{Carlson}, \binits{T.E.}}, \betal:
\batitle{Rectified linear postsynaptic potential function for backpropagation
  in deep spiking neural networks}.
\bjtitle{IEEE transactions on neural networks and learning systems}
\bvolume{33}(\bissue{5}),
\bfpage{1947}--\blpage{1958}
(\byear{2021})
\end{barticle}
\endbibitem

%%% 20
\bibitem[\protect\citeauthoryear{Xu et~al.}{2013}]{xu2013supervised}
\begin{barticle}
\bauthor{\bsnm{Xu}, \binits{Y.}},
\bauthor{\bsnm{Zeng}, \binits{X.}},
\bauthor{\bsnm{Han}, \binits{L.}},
\bauthor{\bsnm{Yang}, \binits{J.}}:
\batitle{A supervised multi-spike learning algorithm based on gradient descent
  for spiking neural networks}.
\bjtitle{Neural Networks}
\bvolume{43},
\bfpage{99}--\blpage{113}
(\byear{2013})
\end{barticle}
\endbibitem

%%% 21
\bibitem[\protect\citeauthoryear{Shrestha and
  Song}{2017}]{shrestha2017robustness}
\begin{barticle}
\bauthor{\bsnm{Shrestha}, \binits{S.B.}},
\bauthor{\bsnm{Song}, \binits{Q.}}:
\batitle{Robustness to training disturbances in spikeprop learning}.
\bjtitle{IEEE transactions on neural networks and learning systems}
\bvolume{29}(\bissue{7}),
\bfpage{3126}--\blpage{3139}
(\byear{2017})
\end{barticle}
\endbibitem

%%% 22
\bibitem[\protect\citeauthoryear{He et~al.}{2016}]{he2016deep}
\begin{bchapter}
\bauthor{\bsnm{He}, \binits{K.}},
\bauthor{\bsnm{Zhang}, \binits{X.}},
\bauthor{\bsnm{Ren}, \binits{S.}},
\bauthor{\bsnm{Sun}, \binits{J.}}:
\bctitle{Deep residual learning for image recognition}.
In: \bbtitle{CVPR},
pp. \bfpage{770}--\blpage{778}
(\byear{2016})
\end{bchapter}
\endbibitem

%%% 23
\bibitem[\protect\citeauthoryear{Ma et~al.}{2018}]{ma2018shufflenet}
\begin{bchapter}
\bauthor{\bsnm{Ma}, \binits{N.}},
\bauthor{\bsnm{Zhang}, \binits{X.}},
\bauthor{\bsnm{Zheng}, \binits{H.-T.}},
\bauthor{\bsnm{Sun}, \binits{J.}}:
\bctitle{Shufflenet v2: Practical guidelines for efficient cnn architecture
  design}.
In: \bbtitle{Proceedings of the European Conference on Computer Vision (ECCV)},
pp. \bfpage{116}--\blpage{131}
(\byear{2018})
\end{bchapter}
\endbibitem

%%% 24
\bibitem[\protect\citeauthoryear{Kahana et~al.}{2022}]{kahana2022physically}
\begin{barticle}
\bauthor{\bsnm{Kahana}, \binits{A.}},
\bauthor{\bsnm{Turkel}, \binits{E.}},
\bauthor{\bsnm{Dekel}, \binits{S.}},
\bauthor{\bsnm{Givoli}, \binits{D.}}:
\batitle{A physically-informed deep-learning model using time-reversal for
  locating a source from sparse and highly noisy sensors data}.
\bjtitle{Journal of Computational Physics}
\bvolume{470},
\bfpage{111592}
(\byear{2022})
\end{barticle}
\endbibitem

%%% 25
\bibitem[\protect\citeauthoryear{Bardos and
  Fink}{2002}]{bardos2002mathematical}
\begin{barticle}
\bauthor{\bsnm{Bardos}, \binits{C.}},
\bauthor{\bsnm{Fink}, \binits{M.}}:
\batitle{Mathematical foundations of the time reversal mirror}.
\bjtitle{Asymptotic Analysis}
\bvolume{29}(\bissue{2}),
\bfpage{157}--\blpage{182}
(\byear{2002})
\end{barticle}
\endbibitem

%%% 26
\bibitem[\protect\citeauthoryear{Givoli and Turkel}{2012}]{givoli2012time}
\begin{barticle}
\bauthor{\bsnm{Givoli}, \binits{D.}},
\bauthor{\bsnm{Turkel}, \binits{E.}}:
\batitle{Time reversal with partial information for wave refocusing and
  scatterer identification}.
\bjtitle{Computer methods in applied mechanics and engineering}
\bvolume{213},
\bfpage{223}--\blpage{242}
(\byear{2012})
\end{barticle}
\endbibitem

%%% 27
\bibitem[\protect\citeauthoryear{Sengupta et~al.}{2019}]{sengupta2019going}
\begin{barticle}
\bauthor{\bsnm{Sengupta}, \binits{A.}},
\bauthor{\bsnm{Ye}, \binits{Y.}},
\bauthor{\bsnm{Wang}, \binits{R.}},
\bauthor{\bsnm{Liu}, \binits{C.}},
\bauthor{\bsnm{Roy}, \binits{K.}}:
\batitle{Going deeper in spiking neural networks: Vgg and residual
  architectures}.
\bjtitle{Frontiers in neuroscience}
\bvolume{13},
\bfpage{95}
(\byear{2019})
\end{barticle}
\endbibitem

%%% 28
\bibitem[\protect\citeauthoryear{Han et~al.}{2020}]{han2020rmp}
\begin{bchapter}
\bauthor{\bsnm{Han}, \binits{B.}},
\bauthor{\bsnm{Srinivasan}, \binits{G.}},
\bauthor{\bsnm{Roy}, \binits{K.}}:
\bctitle{Rmp-snn: Residual membrane potential neuron for enabling deeper
  high-accuracy and low-latency spiking neural network}.
In: \bbtitle{Proceedings of the IEEE/CVF Conference on Computer Vision and
  Pattern Recognition},
pp. \bfpage{13558}--\blpage{13567}
(\byear{2020})
\end{bchapter}
\endbibitem

%%% 29
\bibitem[\protect\citeauthoryear{Diehl et~al.}{2015}]{diehl2015fast}
\begin{bchapter}
\bauthor{\bsnm{Diehl}, \binits{P.U.}},
\bauthor{\bsnm{Neil}, \binits{D.}},
\bauthor{\bsnm{Binas}, \binits{J.}},
\bauthor{\bsnm{Cook}, \binits{M.}},
\bauthor{\bsnm{Liu}, \binits{S.-C.}},
\bauthor{\bsnm{Pfeiffer}, \binits{M.}}:
\bctitle{Fast-classifying, high-accuracy spiking deep networks through weight
  and threshold balancing}.
In: \bbtitle{2015 International Joint Conference on Neural Networks (IJCNN)},
pp. \bfpage{1}--\blpage{8}
(\byear{2015}).
\bcomment{ieee}
\end{bchapter}
\endbibitem

%%% 30
\bibitem[\protect\citeauthoryear{Rueckauer
  et~al.}{2017}]{rueckauer2017conversion}
\begin{barticle}
\bauthor{\bsnm{Rueckauer}, \binits{B.}},
\bauthor{\bsnm{Lungu}, \binits{I.-A.}},
\bauthor{\bsnm{Hu}, \binits{Y.}},
\bauthor{\bsnm{Pfeiffer}, \binits{M.}},
\bauthor{\bsnm{Liu}, \binits{S.-C.}}:
\batitle{Conversion of continuous-valued deep networks to efficient
  event-driven networks for image classification}.
\bjtitle{Frontiers in neuroscience}
\bvolume{11},
\bfpage{682}
(\byear{2017})
\end{barticle}
\endbibitem

%%% 31
\bibitem[\protect\citeauthoryear{Li et~al.}{2021}]{li2021free}
\begin{botherref}
\oauthor{\bsnm{Li}, \binits{Y.}},
\oauthor{\bsnm{Deng}, \binits{S.}},
\oauthor{\bsnm{Dong}, \binits{X.}},
\oauthor{\bsnm{Gong}, \binits{R.}},
\oauthor{\bsnm{Gu}, \binits{S.}}:
A free lunch from ann: Towards efficient, accurate spiking neural networks
  calibration.
arXiv preprint arXiv:2106.06984
(2021)
\end{botherref}
\endbibitem

%%% 32
\bibitem[\protect\citeauthoryear{Lee et~al.}{2016}]{lee2016training}
\begin{barticle}
\bauthor{\bsnm{Lee}, \binits{J.H.}},
\bauthor{\bsnm{Delbruck}, \binits{T.}},
\bauthor{\bsnm{Pfeiffer}, \binits{M.}}:
\batitle{Training deep spiking neural networks using backpropagation}.
\bjtitle{Frontiers in neuroscience}
\bvolume{10},
\bfpage{508}
(\byear{2016})
\end{barticle}
\endbibitem

%%% 33
\bibitem[\protect\citeauthoryear{Neftci et~al.}{2019}]{neftci2019surrogate}
\begin{barticle}
\bauthor{\bsnm{Neftci}, \binits{E.O.}},
\bauthor{\bsnm{Mostafa}, \binits{H.}},
\bauthor{\bsnm{Zenke}, \binits{F.}}:
\batitle{Surrogate gradient learning in spiking neural networks}.
\bjtitle{IEEE Signal Processing Magazine}
\bvolume{36},
\bfpage{61}--\blpage{63}
(\byear{2019})
\end{barticle}
\endbibitem

%%% 34
\bibitem[\protect\citeauthoryear{Shrestha and
  Orchard}{2018}]{shrestha2018slayer}
\begin{botherref}
\oauthor{\bsnm{Shrestha}, \binits{S.B.}},
\oauthor{\bsnm{Orchard}, \binits{G.}}:
Slayer: Spike layer error reassignment in time.
arXiv preprint arXiv:1810.08646
(2018)
\end{botherref}
\endbibitem

%%% 35
\bibitem[\protect\citeauthoryear{Wu et~al.}{2018}]{wu2018spatio}
\begin{barticle}
\bauthor{\bsnm{Wu}, \binits{Y.}},
\bauthor{\bsnm{Deng}, \binits{L.}},
\bauthor{\bsnm{Li}, \binits{G.}},
\bauthor{\bsnm{Zhu}, \binits{J.}},
\bauthor{\bsnm{Shi}, \binits{L.}}:
\batitle{Spatio-temporal backpropagation for training high-performance spiking
  neural networks}.
\bjtitle{Frontiers in neuroscience}
\bvolume{12},
\bfpage{331}
(\byear{2018})
\end{barticle}
\endbibitem

%%% 36
\bibitem[\protect\citeauthoryear{Wu et~al.}{2021}]{wu2021training}
\begin{botherref}
\oauthor{\bsnm{Wu}, \binits{H.}},
\oauthor{\bsnm{Zhang}, \binits{Y.}},
\oauthor{\bsnm{Weng}, \binits{W.}},
\oauthor{\bsnm{Zhang}, \binits{Y.}},
\oauthor{\bsnm{Xiong}, \binits{Z.}},
\oauthor{\bsnm{Zha}, \binits{Z.-J.}},
\oauthor{\bsnm{Sun}, \binits{X.}},
\oauthor{\bsnm{Wu}, \binits{F.}}:
Training spiking neural networks with accumulated spiking flow.
ijo
\textbf{1}(1)
(2021)
\end{botherref}
\endbibitem

%%% 37
\bibitem[\protect\citeauthoryear{Li et~al.}{2021}]{li2021differentiable}
\begin{barticle}
\bauthor{\bsnm{Li}, \binits{Y.}},
\bauthor{\bsnm{Guo}, \binits{Y.}},
\bauthor{\bsnm{Zhang}, \binits{S.}},
\bauthor{\bsnm{Deng}, \binits{S.}},
\bauthor{\bsnm{Hai}, \binits{Y.}},
\bauthor{\bsnm{Gu}, \binits{S.}}:
\batitle{Differentiable spike: Rethinking gradient-descent for training spiking
  neural networks}.
\bjtitle{Advances in Neural Information Processing Systems}
\bvolume{34},
\bfpage{23426}--\blpage{23439}
(\byear{2021})
\end{barticle}
\endbibitem

%%% 38
\bibitem[\protect\citeauthoryear{Kim et~al.}{2022}]{kim2022neural}
\begin{bchapter}
\bauthor{\bsnm{Kim}, \binits{Y.}},
\bauthor{\bsnm{Li}, \binits{Y.}},
\bauthor{\bsnm{Park}, \binits{H.}},
\bauthor{\bsnm{Venkatesha}, \binits{Y.}},
\bauthor{\bsnm{Panda}, \binits{P.}}:
\bctitle{Neural architecture search for spiking neural networks}.
In: \bbtitle{Computer Vision--ECCV 2022: 17th European Conference, Tel Aviv,
  Israel, October 23--27, 2022, Proceedings, Part XXIV},
pp. \bfpage{36}--\blpage{56}
(\byear{2022}).
\bcomment{Springer}
\end{bchapter}
\endbibitem

%%% 39
\bibitem[\protect\citeauthoryear{Wu et~al.}{2020}]{wu2020progressive}
\begin{botherref}
\oauthor{\bsnm{Wu}, \binits{J.}},
\oauthor{\bsnm{Xu}, \binits{C.}},
\oauthor{\bsnm{Zhou}, \binits{D.}},
\oauthor{\bsnm{Li}, \binits{H.}},
\oauthor{\bsnm{Tan}, \binits{K.C.}}:
Progressive tandem learning for pattern recognition with deep spiking neural
  networks.
arXiv preprint arXiv:2007.01204
(2020)
\end{botherref}
\endbibitem

%%% 40
\bibitem[\protect\citeauthoryear{Srivastava
  et~al.}{2015}]{srivastava2015training}
\begin{botherref}
\oauthor{\bsnm{Srivastava}, \binits{R.K.}},
\oauthor{\bsnm{Greff}, \binits{K.}},
\oauthor{\bsnm{Schmidhuber}, \binits{J.}}:
Training very deep networks.
Advances in neural information processing systems
\textbf{28}
(2015)
\end{botherref}
\endbibitem

%%% 41
\bibitem[\protect\citeauthoryear{Huang et~al.}{2017}]{huang2017densely}
\begin{bchapter}
\bauthor{\bsnm{Huang}, \binits{G.}},
\bauthor{\bsnm{Liu}, \binits{Z.}},
\bauthor{\bsnm{Van Der~Maaten}, \binits{L.}},
\bauthor{\bsnm{Weinberger}, \binits{K.Q.}}:
\bctitle{Densely connected convolutional networks}.
In: \bbtitle{Proceedings of the IEEE Conference on Computer Vision and Pattern
  Recognition},
pp. \bfpage{4700}--\blpage{4708}
(\byear{2017})
\end{bchapter}
\endbibitem

%%% 42
\bibitem[\protect\citeauthoryear{Weisstein}{2002}]{weisstein2002z}
\begin{botherref}
\oauthor{\bsnm{Weisstein}, \binits{E.W.}}:
Z-transform.
https://mathworld. wolfram. com/
(2002)
\end{botherref}
\endbibitem

%%% 43
\bibitem[\protect\citeauthoryear{Kim et~al.}{2022}]{kim2022rate}
\begin{bchapter}
\bauthor{\bsnm{Kim}, \binits{Y.}},
\bauthor{\bsnm{Park}, \binits{H.}},
\bauthor{\bsnm{Moitra}, \binits{A.}},
\bauthor{\bsnm{Bhattacharjee}, \binits{A.}},
\bauthor{\bsnm{Venkatesha}, \binits{Y.}},
\bauthor{\bsnm{Panda}, \binits{P.}}:
\bctitle{Rate coding or direct coding: Which one is better for accurate,
  robust, and energy-efficient spiking neural networks?}
In: \bbtitle{ICASSP 2022-2022 IEEE International Conference on Acoustics,
  Speech and Signal Processing (ICASSP)},
pp. \bfpage{71}--\blpage{75}
(\byear{2022}).
\bcomment{IEEE}
\end{bchapter}
\endbibitem

%%% 44
\bibitem[\protect\citeauthoryear{LeCun}{1998}]{lecun1998mnist}
\begin{botherref}
\oauthor{\bsnm{LeCun}, \binits{Y.}}:
The mnist database of handwritten digits.
http://yann. lecun. com/exdb/mnist/
(1998)
\end{botherref}
\endbibitem

%%% 45
\bibitem[\protect\citeauthoryear{Xiao et~al.}{2017}]{xiao2017fashion}
\begin{botherref}
\oauthor{\bsnm{Xiao}, \binits{H.}},
\oauthor{\bsnm{Rasul}, \binits{K.}},
\oauthor{\bsnm{Vollgraf}, \binits{R.}}:
Fashion-mnist: a novel image dataset for benchmarking machine learning
  algorithms.
arXiv preprint arXiv:1708.07747
(2017)
\end{botherref}
\endbibitem

%%% 46
\bibitem[\protect\citeauthoryear{Loshchilov and
  Hutter}{2016}]{loshchilov2016sgdr}
\begin{botherref}
\oauthor{\bsnm{Loshchilov}, \binits{I.}},
\oauthor{\bsnm{Hutter}, \binits{F.}}:
Sgdr: Stochastic gradient descent with warm restarts.
arXiv preprint arXiv:1608.03983
(2016)
\end{botherref}
\endbibitem

%%% 47
\bibitem[\protect\citeauthoryear{Kheradpisheh and
  Masquelier}{2020}]{kheradpisheh2020temporal}
\begin{barticle}
\bauthor{\bsnm{Kheradpisheh}, \binits{S.R.}},
\bauthor{\bsnm{Masquelier}, \binits{T.}}:
\batitle{Temporal backpropagation for spiking neural networks with one spike
  per neuron}.
\bjtitle{International journal of neural systems}
\bvolume{30}(\bissue{06}),
\bfpage{2050027}
(\byear{2020})
\end{barticle}
\endbibitem

%%% 48
\bibitem[\protect\citeauthoryear{Kheradpisheh
  et~al.}{2022}]{kheradpisheh2022bs4nn}
\begin{barticle}
\bauthor{\bsnm{Kheradpisheh}, \binits{S.R.}},
\bauthor{\bsnm{Mirsadeghi}, \binits{M.}},
\bauthor{\bsnm{Masquelier}, \binits{T.}}:
\batitle{Bs4nn: binarized spiking neural networks with temporal coding and
  learning}.
\bjtitle{Neural Processing Letters}
\bvolume{54}(\bissue{2}),
\bfpage{1255}--\blpage{1273}
(\byear{2022})
\end{barticle}
\endbibitem

%%% 49
\bibitem[\protect\citeauthoryear{Sakemi et~al.}{2021}]{sakemi2021supervised}
\begin{botherref}
\oauthor{\bsnm{Sakemi}, \binits{Y.}},
\oauthor{\bsnm{Morino}, \binits{K.}},
\oauthor{\bsnm{Morie}, \binits{T.}},
\oauthor{\bsnm{Aihara}, \binits{K.}}:
A supervised learning algorithm for multilayer spiking neural networks based on
  temporal coding toward energy-efficient vlsi processor design.
IEEE Transactions on Neural Networks and Learning Systems
(2021)
\end{botherref}
\endbibitem

%%% 50
\bibitem[\protect\citeauthoryear{Hao et~al.}{2020}]{hao2020biologically}
\begin{barticle}
\bauthor{\bsnm{Hao}, \binits{Y.}},
\bauthor{\bsnm{Huang}, \binits{X.}},
\bauthor{\bsnm{Dong}, \binits{M.}},
\bauthor{\bsnm{Xu}, \binits{B.}}:
\batitle{A biologically plausible supervised learning method for spiking neural
  networks using the symmetric stdp rule}.
\bjtitle{Neural Networks}
\bvolume{121},
\bfpage{387}--\blpage{395}
(\byear{2020})
\end{barticle}
\endbibitem

%%% 51
\bibitem[\protect\citeauthoryear{Ranjan et~al.}{2020}]{ranjan2020novel}
\begin{barticle}
\bauthor{\bsnm{Ranjan}, \binits{J.A.K.}},
\bauthor{\bsnm{Sigamani}, \binits{T.}},
\bauthor{\bsnm{Barnabas}, \binits{J.}}:
\batitle{A novel and efficient classifier using spiking neural network}.
\bjtitle{The Journal of Supercomputing}
\bvolume{76}(\bissue{9}),
\bfpage{6545}--\blpage{6560}
(\byear{2020})
\end{barticle}
\endbibitem

%%% 52
\bibitem[\protect\citeauthoryear{Park et~al.}{2020}]{park2020t2fsnn}
\begin{botherref}
\oauthor{\bsnm{Park}, \binits{S.}},
\oauthor{\bsnm{Kim}, \binits{S.}},
\oauthor{\bsnm{Na}, \binits{B.}},
\oauthor{\bsnm{Yoon}, \binits{S.}}:
T2fsnn: Deep spiking neural networks with time-to-first-spike coding.
arXiv preprint arXiv:2003.11741
(2020)
\end{botherref}
\endbibitem

%%% 53
\bibitem[\protect\citeauthoryear{Horowitz}{2014}]{horowitz20141}
\begin{bchapter}
\bauthor{\bsnm{Horowitz}, \binits{M.}}:
\bctitle{1.1 computing's energy problem (and what we can do about it)}.
In: \bbtitle{2014 IEEE International Solid-State Circuits Conference Digest of
  Technical Papers (ISSCC)},
pp. \bfpage{10}--\blpage{14}
(\byear{2014}).
\bcomment{IEEE}
\end{bchapter}
\endbibitem

%%% 54
\bibitem[\protect\citeauthoryear{Grumiaux et~al.}{2022}]{Grumiaux_2022}
\begin{barticle}
\bauthor{\bsnm{Grumiaux}, \binits{P.-A.}},
\bauthor{\bsnm{Kiti{\'{c}}}, \binits{S.}},
\bauthor{\bsnm{Girin}, \binits{L.}},
\bauthor{\bsnm{Gu{\'{e}}rin}, \binits{A.}}:
\batitle{A survey of sound source localization with deep learning methods}.
\bjtitle{The Journal of the Acoustical Society of America}
\bvolume{152}(\bissue{1}),
\bfpage{107}--\blpage{151}
(\byear{2022})
\doiurl{10.1121/10.0011809}
\end{barticle}
\endbibitem

%%% 55
\bibitem[\protect\citeauthoryear{Ovadia et~al.}{2023}]{ovadia2023vito}
\begin{botherref}
\oauthor{\bsnm{Ovadia}, \binits{O.}},
\oauthor{\bsnm{Kahana}, \binits{A.}},
\oauthor{\bsnm{Stinis}, \binits{P.}},
\oauthor{\bsnm{Turkel}, \binits{E.}},
\oauthor{\bsnm{Karniadakis}, \binits{G.E.}}:
Vito: Vision transformer-operator
(2023)
{\href{https://arxiv.org/abs/2303.08891}{{arXiv:2303.08891}}}
{[cs.CV]}
\end{botherref}
\endbibitem

\end{thebibliography}

\end{document}